%% file: ANA-HMBS-2024-46-PAPER.tex
\documentclass[PAPER, cernpreprint, atlasdraft=false, UKenglish, texmf, orcidlogo]{atlasdoc}
\usepackage{atlaspackage}
\usepackage{atlasbiblatex}
\usepackage{multirow}
\usepackage{amsmath} %
\usepackage{siunitx} %

\usepackage{atlasphysics}
\ifthenelse{\boolean{AtlasDraft}}
\graphicspath{{logos/}{figures/}}

\usepackage{ANA-HMBS-2024-46-PAPER-defs}

\usepackage{verbatim}  %
\usepackage{makecell}  %
\usepackage{subfig}

\AtlasTitle{Search for a new pseudoscalar decaying into a pair of bottom and antibottom quarks in top-associated production in $\sqrt{s}=13$~TeV proton--proton collisions with the ATLAS detector}

\AtlasAbstract{%
A search for a pseudoscalar $a$ produced in association with a top-quark pair, or in association with a single top quark plus a $W$ boson, with the pseudoscalar decaying into $b$-quarks ($a\rightarrow b\bar{b}$), is performed using the full Run 2 data sample using a dileptonic decay mode signature.
The search covers pseudoscalar boson masses between $12\-- 100$~GeV and involves both the kinematic regime where the decay products of the pseudoscalar are reconstructed as two standard $b$-tagged small-radius jets, or merged into a large-radius jet due to its Lorentz boost. No significant excess relative to expectations is observed.
Assuming a branching ratio $\text{BR}(a\rightarrow b\bar{b})=100\% $, the range of pseudoscalar masses between 50 and 80~GeV is excluded at 95\% confidence level for a coupling of the pseudoscalar to the top quark of 0.5, while a coupling of 1.0 is excluded at 95\% confidence level for the masses considered, with the coupling defined as the strength modifier of the Standard Model Yukawa coupling.
}

\AtlasRefCode{HMBS-2024-46}

\PreprintIdNumber{CERN-EP-2025-060}

\AtlasJournal{\EPJC}
\AtlasJournalRef{\EPJC 85, 886 (2025)}
\AtlasDOI{10.1140/epjc/s10052-025-14507-0}

\hypersetup{pdftitle={ATLAS document},pdfauthor={The ATLAS Collaboration}}

\begin{document}

\maketitle

\tableofcontents

\renewcommand{\arraystretch}{1.2}

\clearpage

\section{Introduction}
\label{sec:intro}

Since the discovery of the Higgs boson~\cite{HIGG-2012-27, CMS-HIG-12-028}, there is an ongoing effort at the Large Hadron Collider (LHC)~\cite{Evans:2008zzb} to measure its properties and search for new physics. The Higgs boson was discovered by observing few decay modes~\cite{HIGG-2012-27, CMS-HIG-12-028} consistent with the Standard Model (SM) predictions. Part of the interest is about the nature of the discovered particle and whether it is the single boson predicted by the SM or, alternatively, part of an extended Higgs sector as suggested by models such as the two-Higgs-doublet model (2HDM)~\cite{Branco_2012, Aoki_2009}, which can be embedded into supersymmetric models~\cite{Golfand:1971iw,Volkov:1973ix,WESS197439,Wess:1974jb,Ferrara:1974pu,Salam:1974ig}.
It predicts a total of five bosons: a light ($h$) and a heavy ($H$) CP-even Higgs boson, with the light one corresponding to the observed Higgs boson; two charged Higgs bosons ($H^{+}$ and $H^{-}$); and a CP-odd particle ($a$), also referred to as pseudoscalar. The additional scalar/pseudoscalar states of these models may also provide a portal into dark matter, serving as a mediator between the SM and dark matter sector~\cite{Buckley_2015,Bauer_2017}. A pseudoscalar
$a$ is also be predicted in axion models~\cite{Kim:1986ax}.

\begin{figure}[h]
\centering
\subfloat[]{\includegraphics[width=0.5\textwidth]{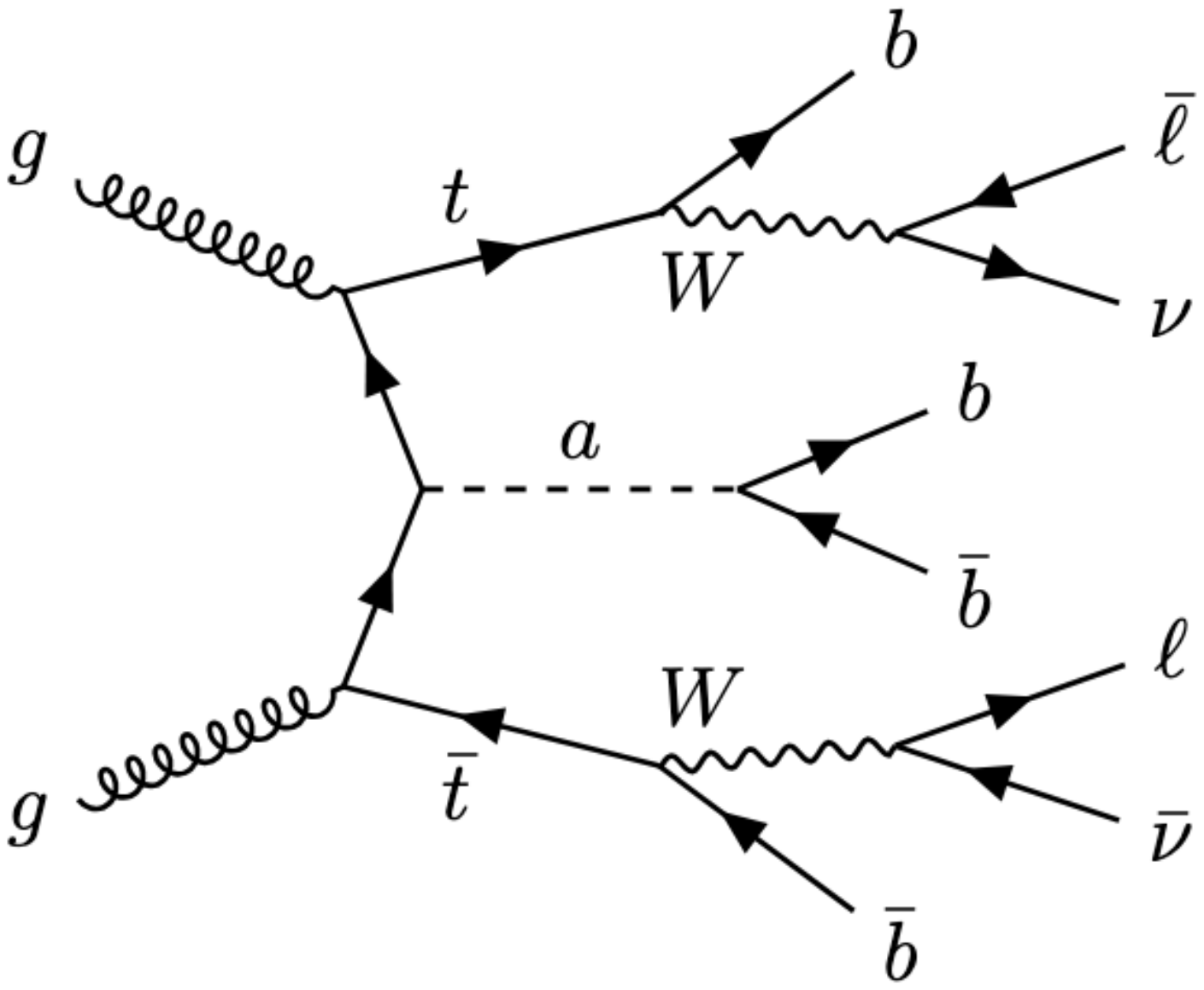}}
\subfloat[]{\includegraphics[width=0.5\textwidth]{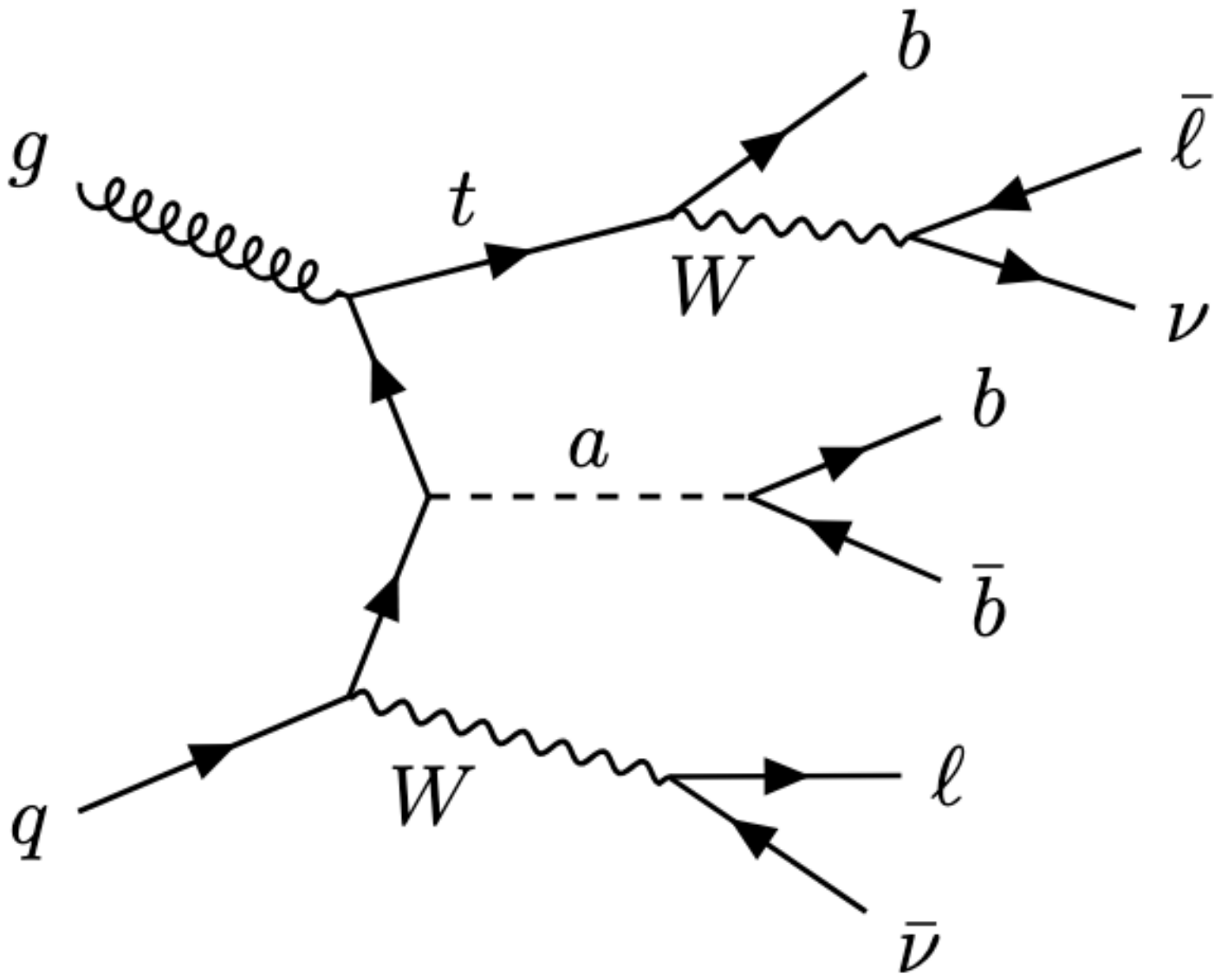}}
\caption{Feynman diagrams for (a) $\ttbar$- and (b) $tW$-associated  production of a pseudoscalar particle $a$ that decays into a pair of $b$-quarks.}
\label{fig:tta_feynman}
\end{figure}

This analysis performs a search for a light pseudoscalar with a mass smaller than the SM Higgs boson, produced in association either with a top-quark pair or a single top quark and a $W$ boson, where the pseudoscalar decays into a bottom-antibottom quark pair, as proposed in~\cite{casolino2015probinglightcpoddscalar}. It is based on a simplified model with the following Yukawa lagrangian:
\begin{equation}
\mathcal{L} = -\frac{g_t y_t}{\sqrt{2}}a \bar{t} (i\gamma^5) t -\frac{g_by_b}{\sqrt{2}}a \bar{b} (i\gamma^5) b
\label{eq:tta_simplified_yukawa},
\end{equation}
where $y_j/\sqrt{2}=m_j/v$ is the SM Yukawa coupling of particle $j$ to the pseudoscalar $a$ and $g_j$ is the coupling modifier, with $j= t$ or $b$.
Figure~\ref{fig:tta_feynman} shows two example Feynman diagrams for this process. This decay channel is favoured by many models for the range of explored pseudoscalar masses, $m_{a} < m_{h}$, although the branching ratios of the pseudoscalar depend on the specific model parameters.
This is the first search for this process, exploring the couplings of the pseudoscalar to bottom quarks. Previously, the ATLAS and CMS Collaborations performed similar searches of \tta associated production exploiting leptonic decays of the pseudoscalar. The CMS Collaboration studied the decay of a pseudoscalar to the three families of leptons~\cite{CMS-EXO-21-018}, while the ATLAS Collaboration studied the pseudoscalar coupling to muons~\cite{HDBS-2020-12}. None of these searches found significant excesses, but these decay channels are typically disfavoured  compared with the $b\bar{b}$ decays when assuming a Yukawa coupling.

The search is performed in the dileptonic decay channel, with both top and antitop quarks (or both $W$ bosons) decaying leptonically. Despite the reduced branching ratio of this decay channel, the reduced jet multiplicity of the final state and the precisely measured kinematics of the two leptons allow for a more efficient identification of the $b$-jets originating from the top and antitop quark decays than its semileptonic (with only one $W$ boson from either the top or antitop quarks decaying leptonically) or fully hadronic (with the two $W$ bosons decaying hadronically) counterparts.

For masses of the pseudoscalar below $\sim$30~\GeV, the $b\bar{b}$ pair has a large Lorentz-boost and is thus reconstructed as a single large-radius jet. On the other hand, for higher masses, the jets are well separated at detector level. The analysis is designed to exploit both kinematic regimes to have good signal sensitivity, making use of multiple signal regions, reconstructed objects, and machine learning techniques.

\section{ATLAS detector}
\label{sec:detector}
\newcommand{\AtlasCoordFootnote}{%
ATLAS uses a right-handed coordinate system with its origin at the nominal interaction point (IP) in the centre of the detector and the \(z\)-axis along the beam pipe. The \(x\)-axis points from the IP to the centre of the LHC ring, and the \(y\)-axis points upwards. Polar coordinates \((r,\phi)\) are used in the transverse plane, \(\phi\) being the azimuthal angle around the \(z\)-axis. The pseudorapidity is defined in terms of the polar angle \(\theta\) as \(\eta = -\ln \tan(\theta/2)\) and is equal to the rapidity $y=\frac{1}{2}\ln\left(\frac{E+p_{z}}{E-p_{z}}\right)$ in the relativistic limit. Angular distance is measured in units of \smash{\(\Delta R \equiv \sqrt{(\Delta y)^{2} + (\Delta\phi)^{2}}\)}.}

The ATLAS detector~\cite{PERF-2007-01} at the LHC covers nearly the entire solid angle around the collision point.\footnote{\AtlasCoordFootnote}
It consists of an inner tracking detector surrounded by a thin superconducting solenoid, electromagnetic and hadronic calorimeters,
and a muon spectrometer incorporating three large superconducting air-core toroidal magnetic systems.

The inner-detector system (ID) is immersed in a \qty{2}{\tesla} axial magnetic field
and provides charged-particle tracking in the range of \(|\eta| < 2.5\).
The high-granularity silicon pixel detector covers the vertex region and typically provides four measurements per track,
the first hit generally being in the insertable B-layer installed before Run~2~\cite{ATLAS-TDR-19,PIX-2018-001}.
It is followed by the SemiConductor Tracker, which usually provides eight measurements per track.
These silicon detectors are complemented by the transition radiation tracker (TRT),
which enables radially extended track reconstruction up to \(|\eta| = 2.0\).
The TRT also provides electron identification information
based on the fraction of hits (typically 30 in total) above a higher energy-deposit threshold corresponding to transition radiation.

The calorimeter system covers the pseudorapidity range \(|\eta| < 4.9\).
Within the region \(|\eta|< 3.2\), electromagnetic calorimetry is provided by barrel and
endcap high-granularity lead/liquid-argon (LAr) calorimeters,
with an additional thin LAr presampler covering \(|\eta| < 1.8\)
to correct for energy loss in material upstream of the calorimeters.
Hadronic calorimetry is provided by the steel/scintillator-tile calorimeter,
segmented into three barrel structures within \(|\eta| < 1.7\), and two copper/LAr hadronic endcap calorimeters.
The solid angle coverage is completed with forward copper/LAr and tungsten/LAr calorimeter modules
optimised for electromagnetic and hadronic energy measurements respectively.

The muon spectrometer comprises separate trigger and
high-precision tracking chambers measuring the deflection of muons in a magnetic field generated by the superconducting air-core toroidal magnets.
The field integral of the toroids ranges between \num{2.0} and \qty{6.0}{\tesla\metre}
across most of the detector.
Three layers of precision chambers, each consisting of layers of monitored drift tubes, cover the region \(|\eta| < 2.7\),
complemented by cathode-strip chambers in the forward region, where the background is highest.
The muon trigger system covers the range \(|\eta| < 2.4\) with resistive-plate chambers in the barrel, and thin-gap chambers in the endcap regions.

The luminosity is measured mainly by the LUCID--2~\cite{LUCID2} detector that records Cherenkov light produced in the quartz windows of photomultipliers located close to the beampipe.

Events are selected by the first-level trigger system implemented in custom hardware,
followed by selections made by algorithms implemented in software in the high-level trigger~\cite{TRIG-2016-01}.
The first-level trigger accepts events from the \qty{40}{\MHz} bunch crossings at a rate below \qty{100}{\kHz},
which the high-level trigger further reduces to record complete events to disk at about \qty{1}{\kHz}.

A software suite~\cite{SOFT-2022-02} is used in data simulation, in the reconstruction
and analysis of real and simulated data, in detector operations, and in the trigger and data acquisition
systems of the experiment.

\nopagebreak


\section{Data and simulated event samples}
\label{ch:data_and_simulated_samples}

This search is based on proton--proton ($pp$) collision data at a centre-of-mass energy of 13~\TeV collected with the ATLAS detector at the LHC from 2015 to 2018, referred to as the Run~2 data sample in the following. After applying the data quality requirements that ensure that all subdetectors were operational, the integrated luminosity of the data sample is $140.1\pm1.2$~\ifb\cite{DAPR-2021-01}.

The signal consists in the production of a light pseudoscalar $a$ in association with a top-quark pair, \tta, or in association with a single top quark and a $W$ boson, $tWa$. The main background in this search is \ttbar production in association with jets, followed by smaller contributions from single top quark, \ttH, \ttV, $V$+jets, diboson and other rare processes involving the production of a top quark. The analysis only considers the decay of the pseudoscalar to a $b\Bar{b}$~pair. Decays of the pseudoscalar to other final states like a $\tau\Bar{\tau}$~pair or a $c\Bar{c}$~pair are not considered, as these are suppressed both by the Yukawa coupling for the masses considered (roughly a factor 5 and 20, respectively) and by the large $b$-jet multiplicity required in the analysis signal regions.

All signal and background samples are simulated using various Monte Carlo (MC) matrix-element (ME) generators interfaced with different algorithms for the parton shower, hadronisation, and underlying event.
The effect from multiple $pp$ interactions originating from the same or neighbouring bunch crossings, usually referred to as pile-up, is simulated
by overlaying the simulated hard-scattering event with inelastic $pp$ collisions simulated using \PYTHIA[8.1]~\cite{Sjostrand:2007gs} and the A3 set of tuned parameters (tune) ~\cite{ATL-PHYS-PUB-2016-017}.
A reweighting is applied to the simulated samples such that they match the pile-up conditions in data.
For the detector simulation two different approaches are used. The full ATLAS detector simulation (FS) is based on
\textsc{Geant}~4~\cite{Agostinelli:2002hh}, while the ``fast'' detector simulation (AF2) uses a parameterisation of the calorimeter response~\cite{SOFT-2010-01}. Most background samples are produced with FS while signal samples are produced with AF2. Both MC and data are processed using the same reconstruction and analysis software.

The $t\Bar{t}a$ signal samples are simulated with \MGNLO[2.3.3]~\cite{Alwall:2014hca} generator at next-to-leading order~(NLO) in the strong coupling constant \alphas. A simplified model based on the decoupling limit of the 2HDM+a type II is used. Additionally, the subdominant $tWa$ signal samples are simulated with the same generator at leading-order~(LO). For notational simplicity, in the following \tta refers to the production of the pseudoscalar $a$ in association with either a top-quark pair or a single top quark and a $W$ boson.
Samples are simulated for the following values of the pseudoscalar mass $m_{a}$: 12, 16, 20, 25, 30, 40, 50, 60, 80 and 100~\GeV.
Additional $t\Bar{t}a$ signal samples for 20 and 60~\GeV are also simulated with FS to check that no significant differences between the two detector simulations are observed.

The production of top-quark pairs with additional jets represents the main background source, especially the production of \ttbar plus heavy flavour (\ttbar+HF): $t\bar{t}+b$-jets and $t\bar{t}+c$-jets. For the modelling of $t\bar{t}+b$-jets events, four flavour-scheme samples (4FS), with massive $b$-quarks, are simulated using the \textsc{PowhegBox}-Res framework at NLO~\cite{ATL-PHYS-PUB-2022-006} and the \NNPDF[3.1nnlo] parton distribution function (PDF) set is used. For the modelling of $t\bar{t}+c$-jets and $t\bar{t}+$light-jets events, five flavour-scheme (5FS) samples with massless $b$-quarks are simulated using \textsc{PowhegBox}-v2~\cite{Frixione:2007nw,Nason:2004rx,Frixione:2007vw,Alioli:2010xd}, also at NLO\@. To avoid double-counting of events, neither the $t\bar{t}+b$-jets events from the 5FS samples nor the $t\bar{t}+c$-jets nor the $t\bar{t}+$light-jets events from the 4FS samples are used. The simulated $t\bar{t}$ events are categorised based on the number of additional jets matched to $b$- or $c$-hadrons with transverse momentum \pT\ larger than 5~\GeV within $\Delta R<0.3$ of the jet axis.

To model the production of single-top-quark events, which mostly contribute through the $tW$-channel, and the production of $t\bar{t}H$, the same \textsc{PowhegBox}-v2~\cite{Frixione:2007nw,Nason:2004rx,Frixione:2007vw,Alioli:2010xd} settings, as used in the $t\bar{t}+$light/$c$-jets production, are used.
The production of \ttZ and \ttW events is modelled using the
\MGNLO[2.3.3]~\cite{Alwall:2014hca} generator at NLO\@.
For all samples listed above, the \NNPDF[3.0nlo]~\cite{Ball:2014uwa} PDF sets are used and a top-quark mass of \(\mtop = \SI{172.5}{\GeV}\) is set. The events are interfaced with \PYTHIA[8.230]~\cite{Sjostrand:2014zea}~using
the A14 tune~\cite{ATL-PHYS-PUB-2014-021} and the \NNPDF[2.3lo]
set of PDFs~\cite{Ball:2012cx} for the parton shower and hadronisation modelling.

The production of $tZq$ and $tWZ$ events is performed using \textsc{MadGraph}5\_aMC@NLO~v2.3.3, at LO and NLO in QCD respectively, in the 4FS with the \texttt{CTEQ6L1} PDF set~\cite{Pumplin:2002vw} and using \textsc{Pythia}~8.212 for the parton shower.

Finally, the production of \(V+\)jets and diboson samples (\(VV\)) is simulated with  different versions of the \SHERPA~\cite{Bothmann:2019yzt} generator and the simulated events are matched with the \SHERPA parton shower~\cite{Schumann:2007mg} using the \MEPSatNLO prescription~\cite{Hoeche:2011fd,Hoeche:2012yf,Catani:2001cc,Hoeche:2009rj} with the set of tuned parameters developed by the \SHERPA authors. All samples and their basic generation parameters are summarised in Table~\ref{tab:MC_samples}.

For all samples, except those generated with \SHERPA, the decays of $b$- and $c$-hadrons are simulated using the \EVTGEN programme~\cite{Lange:2001uf}.

\begin{table}[h]
\caption{Nominal simulated signal and background event samples. The matrix element generator, PDF set, parton-shower (PS) generator and calculation
accuracy of the cross section in QCD and EW used for normalisation are shown. \textsc{MadGraph} is abbreviated to MG.}
\footnotesize
\centering
\renewcommand{\arraystretch}{1.4}
\hspace*{-0.4cm}
\begin{tabular}{l|l|l|l|l}

Process & Matrix Element generator & PDF set & PS generator & Normalisation  \\ \hline \hline
$t\bar{t}a$, $a\rightarrow b\bar{b}$ &
\multirow{2}{*}{MG\_aMC@NLO v2.3.3} &
\multirow{2}{*}{\texttt{NNPDF3.0}~NLO} &
\multirow{2}{*}{\textsc{Pythia}~8.230} &
\multirow{2}{*}{--}  \\
$tWa$, $a\rightarrow b\bar{b}$ & & & & \\ \hline
$t\bar{t}$ + jets &
\textsc{PowhegBox}-v2 &
\texttt{NNPDF3.0}~NLO &
\textsc{Pythia}~8.230 &
(NLO+NNLL)$_\text{QCD}$~\cite{Beneke:2011mq,Cacciari:2011hy,Baernreuther:2012ws,Czakon:2012zr,Czakon:2012pz,Czakon:2013goa,Czakon:2011xx}  \\ \hline
$t\bar{t}$ + $b\bar{b}$ &
\textsc{PowhegBox}-Res &
\texttt{NNPDF3.1}~NNLO  &
\textsc{Pythia}~8.244 &
-- \\ \hline
Single-top &
\textsc{PowhegBox}-v2 &
\texttt{NNPDF3.0}~NLO &
\textsc{Pythia}~8.230 &
(NLO+NNLL)$_\text{QCD}$~\cite{Kidonakis:2010ux,Kidonakis:2013zqa}  \\ \hline
$t\bar{t}H$ &
\textsc{PowhegBox}-v2 &
\texttt{NNPDF3.0}~NLO &
\textsc{Pythia}~8.230 &
NLO$_\text{QCD+EW}$~\cite{deFlorian:2016spz}  \\ \hline
$t\bar{t}Z$ &
MG5\_aMC@NLO v2.3.3 &
\texttt{NNPDF3.0}~NLO &
\textsc{Pythia}~8.210 &
NLO$_\text{QCD+EW}$~\cite{deFlorian:2016spz}  \\[-2pt]
\hline
$t\bar{t}W$ &
MG5\_aMC@NLO v2.3.3&
\texttt{NNPDF3.0}~NLO &
\textsc{Pythia}~8.210 &
NLO$_\text{QCD+EW}$~\cite{deFlorian:2016spz}  \\[-2pt]
\hline
\multirow{2}{*}{$tZq$, $tWZ$} &
\multirow{2}{*}{MG5\_aMC@NLO v2.3.3}&
\texttt{CTEQ6L1} &
\multirow{2}{*}{\textsc{Pythia}~8.212} &
\multirow{2}{*}{--}  \\[-2pt]
&
&
\texttt{NNPDF3.0} NLO &
&
\\ \hline
$Z/W$ + jets &
\textsc{Sherpa}~2.2.11 &
\texttt{NNPDF3.1} NNLO &
\textsc{Sherpa} &
NNLO$_\text{QCD}$~\cite{Anastasiou:2003ds}  \\ \hline
\multirow{2}{*}{Diboson} &
\textsc{Sherpa}~2.2.1 &
\multirow{2}{*}{\texttt{NNPDF3.1} NNLO} &
\multirow{2}{*}{\textsc{Sherpa}} &
--  \\[-2pt]
&
\textsc{Sherpa}~2.2.2 &
&
&  \\

\end{tabular}

\label{tab:MC_samples}
\end{table}


%

\section{Object definition}
\label{ch:object_definition}

In this section, the reconstruction and definition of the physics objects are described, together with the additional corrections applied to each.

\subsection{Physics objects}

Electron candidates are reconstructed from energy deposits (clusters) in the electromagnetic calorimeter associated with reconstructed tracks in the inner detector.
Candidates are selected with $\pT>10$~\GeV and $|\eta|<2.47$, excluding the calorimeter transition region $1.37<|\eta|<1.52$. Electrons satisfy the \texttt{TightLH}~\cite{EGAM-2018-01} likelihood-based identification criterion and are required to match the PromptLeptonTagger working point \texttt{PLVLoose}~\cite{HIGG-2017-02}. They are further required to have $|z_0\sin{\theta}| < 0.5$~mm and a $d_0$ significance $|\frac{d_0}{\sigma(d_0)}|<5$, where the transverse impact parameter ($d_0$) is calculated relative to the beamline, and the longitudinal impact parameter ($z_0$) as the longitudinal distance between the primary vertex and the point where $d_0$ is measured.

Muon candidates are reconstructed from track segments in the various layers of the muon spectrometer and matched with tracks from the inner detector. The final muon candidates are refitted using the complete track information from both detector systems and must have $\pT>10$~\GeV and $|\eta|<2.5$. Muons are required to satisfy the \texttt{Medium} quality requirements and match the PromptLeptonTagger working point \texttt{PLVLoose}~\cite{MUON-2018-03}. Further requirements are $|\frac{d_0}{\sigma(d_0)}|<3$, and $|z_0\sin{\theta}|<0.5$~mm.

Small-$R$ jet candidates are reconstructed by clustering particle flow objects~\cite{PERF-2015-09} using the anti-\kt algorithm~\cite{Cacciari:2008gp,Fastjet} with a radius parameter of $R=0.4$ and a four-momentum recombination scheme. The energy of the jet is corrected to the particle level by the application of a jet energy scale calibration derived from $\sqrt{s}=13$ \TeV data and simulation~\cite{JETM-2018-05}. Baseline jets are required to have $\pT>20$~\GeV and $|\eta| < 2.5$.
For pile-up rejection, jets with \pt $\in [20,60]~\GeV$ and $|\eta|<2.4$ are required to have a jet vertex tagger weight~\cite{PERF-2014-03} larger than 0.5.

The $b$-tagging is the identification of jets that originate from the decay of $b$-hadrons using dedicated algorithms. A deep neural-network, called DL1r~\cite{FTAG-2019-07,FTAG-2018-01,FTAG-2020-08,FTAG-2019-02} is used.
Small-$R$ jets with a DL1r score above a certain threshold are defined as $b$-tagged jets. The pseudo-continuous (PC) $b$-tagging working point is used:
each jet is classified with an integer from one to five depending on how many  calibrated $b$-tagging working points (WP) the jet fulfils. The four calibrated DL1r WPs are 85\%, 77\%, 70\% and 60\%, each corresponding to the approximate average $b$-tagging efficiency in an inclusive \ttbar MC sample. Jets not satisfying any WP are assigned a value of one, and this value is increased by one for every WP that they fulfil. The sum of the PC $b$-tagging over all jets in an event is defined as \texttt{sumPCBTag}. Jets satisfying the 70\% WP are referred to as $b$-jets, while jets satisfying the 85\% WP, but not the 70\% WP, are classified as loose-$b$-jets.

Large-$R$ jet candidates are formed by reclustering the small-$R$ jets and tracks with a larger radius parameter of $R=0.8$ using the anti-\kt algorithm~\cite{Cacciari:2008gp,Fastjet}. The larger radius for track association allows more tracks from the targeted double $b$-hadron decays to be associated with the reclustered jet.
The tracks in and around the small-$R$ jet associated with the reclustered jet through ghost association~\cite{CACCIARI2008119,Matteo_Cacciari_2008} are selected with a loose track selection~\cite{ATL-PHYS-PUB-2020-014}. In this procedure, the \pt of the tracks is set to infinitesimal values, such that the ``ghost'' tracks can then be reclustered with the constituents of the reclustered jets with the appropriate radius parameter. Since the \pt of the tracks is infinitesimally small, they do not influence the reconstruction of the jet, allowing the use of additional tracks that leak outside the small-$R$ jets.

To resolve the substructure within a large-R jet originating from a boosted \Xbbbar decay, which the small-R jet reconstruction fails to completely capture,
additional information is extracted from the large-$R$ jet by reconstructing %
track-subjets inside the large-$R$ jet. The %
track-subjets are derived using the tracks that are ghost associated to each large-$R$ jet as inputs to the exclusive-$k_{T}$ method~\cite{ATL-PHYS-PUB-2017-010}. The selected tracks for a given jet are clustered using the $k_{T}$ algorithm with a radius parameter of $R=0.8$. The clustering stops when there are exactly two track clusters left. These clusters are used as the %
track-subjets associated with a given jet. For signal events, each %
track-subjet should originate from the decay of one $b$-hadron ideally.
The associated large-$R$ jet is required to satisfy $|\eta|<2.0$ to account for their extended radius and the acceptance of the ID\@. Furthermore, each %
track-subjet is required to satisfy $\pt > 5\,\GeV$ where the track-subjet $\pt$ is estimated from the sum of its constituent tracks' four-momenta. The four-momentum of the large-$R$ jet is defined as the sum of the four-momenta of its %
track-subjets.

In addition, secondary vertices (SV) inside the large-$R$ jets are reconstructed to help the identification of $b$-hadrons.
For this purpose, an algorithm
that combines the track-cluster-based low-\pt vertex tagger (TC-LVT)~\cite{FTAG-2023-02} and the multiple-secondary-vertex finder algorithms (MSVF)~\cite{Kostyukhin:685550} is used.
The TC-LVT algorithm was developed for soft $b$-hadron tagging and optimised to reconstruct low-\pt $b$-hadron decays. The clustering algorithm from TC-LVT is used to identify displaced tracks not originating from the primary vertex. The MSVF algorithm is used to identify multiple SVs in the track cluster. The algorithm builds all two-track proto-vertices consistent with displaced tracks that are not compatible with a hadronic material interaction, a photon conversion, or the decay of long-lived light-flavoured hadrons. All displaced tracks reconstructed in the ID are used to build proto-vertices. Proto-vertices define track-to-track relations, since a single track can be associated with more than one proto-vertex. Each set of tracks that are mutually connected to each other forms a secondary vertex. After secondary vertices are formed, tracks not compatible with the vertex are removed, and the ambiguity caused by distant vertices sharing common tracks is resolved. Nearby vertices are also merged by the MSVF algorithm. Finally, reconstructed SVs are required to be $\Delta R$-matched to a large-$R$ jet.
Further details and studies about the large-$R$ jets can be found in Ref.~\cite{IDTR-2021-03}.

In contrast to $b$-tagging, the $B$-tagging is the identification of pairs of $b$-jets that are too close to be resolved and identified individually. For this purpose the \DeXTer tagger is used. It is a double $b$-tagger based on a deep sets neural network (NN) architecture designed to do flavour tagging of merged reconstructed jets~\cite{ATL-PHYS-PUB-2022-042} and uses information of the SVs and jet kinematics. This is done in two transverse momentum ranges: a low \pT range between 20 and 200 \GeV and a high \pT one, above 200~\GeV.
Two working points are defined: the 0--40\% tagging interval is referred to as \texttt{Tight} WP, and the \texttt{Loose} WP is defined by the inclusive 40--60\% tagging interval. A sample of $Z$+jets and \ttbar events is used to measure the \DeXTer efficiency in data, and to derive $B$-tagging and $b$-mistagging rate correction factors for the simulated events. Large-$R$ jets satisfying the \texttt{Tight} WP are referred to as $B$-jets.

The missing transverse momentum \met measures the event momentum imbalance in the transverse plane of the detector. It is defined as the magnitude of the negative vector sum of $\pT$ for all selected and calibrated physics objects in the event, with an extra term added to account for soft energy that is not associated with any of the selected objects. This soft term is calculated from inner detector tracks matched to the primary vertex to make it more resilient to pile-up contamination~\cite{JETM-2020-03}.
The \met computation is based on the momenta of the objects defined previously, and after applying the overlap removal procedure defined in the next section.

\subsection{Corrections to physics objects}

The overlap removal is the procedure followed to prevent double-counting of objects.
First, electrons that share a track with a muon are removed. To prevent double-counting of electron energy deposits as jets, the closest small-$R$ jet within $\Delta R<0.2$ of a selected electron is removed. If the nearest jet surviving that selection is within $\Delta R$ = 0.4 of the electron, the electron is discarded.
To reduce the background from muons from heavy-flavour decays inside jets, muon candidates are required to be separated by $\Delta R>0.4$ from the nearest small-$R$ jet, removing the muon if the jet has at least three associated tracks, and removing the jet otherwise. This avoids an inefficiency for high-energy muons undergoing significant energy loss in the calorimeter.

To avoid double-counting of jets, the jet overlap removal is done as follows. First, every small-$R$ jet is tested to see if it is eligible to be \dexter-tagged. This requires the small-$R$ jet to have $\pt > 20$ \GeV and be isolated, meaning it is the only constituent of its reclustered jet~\cite{Nachman:2014kla} with the \antikt algorithm and radius parameter $R=0.8$.
These jets are then tested by the \dexter tagger: if a small-$R$ jet passes the \texttt{Loose} working point selection, it is defined as a $B$-tagged large-$R$ jet and removed from the small-$R$ jet list.
Otherwise, it is kept as a small-$R$ jet.
Thus, this first step gives two jet lists: the \dexter-tagged large-$R$ jets and the small-$R$ jets, which are either not eligible for \dexter tagging or fail the tagger.
Finally, the jet overlap removal procedure with leptons is repeated for large-$R$ jets using $\Delta R$ = 0.8.

The $\mu$-in-jet \pt correction is the procedure of adding the muons reconstructed inside of a $b$-jet to the four-momentum of the respective $b$-jet. Around 20\% of all $b$-hadron decays produce a low-momentum or soft muon inside of the resulting jet, but those soft muons are removed in the overlap removal as described above. This correction recovers the original energy of those $b$-jets and reduces biases
from the invariant masses calculated with them. %
The soft muons used are required to have $\pt > 4~\GeV$ and $|\eta| < 2.5$, and fulfil the \texttt{Medium} soft muon quality requirement.

In the case of the \dexter-tagged jet, the $\mu$-in-jet \pt correction is carried out as follows. First, the soft muons are matched to %
track-subjets within an angular distance of $\Delta R< 0.3$.
At most the two highest \pt soft muons are taken into account for each %
track-subjet, and any muon is only matched once to the closest subjet. At last, the matched muons are added to the four-momentum of the track-subjet.


%

\section{Event selection}
\label{ch:event_selection}

Only events recorded with a single-electron~\cite{TRIG-2018-05} or single-muon trigger~\cite{TRIG-2018-01} under stable beam conditions and for which all detector subsystems were operational are considered~\cite{DAPR-2018-01}. Single-lepton triggers with $\pT$ thresholds varying from 20 to 140~\GeV, depending on the year, lepton flavour, isolation requirement and luminosity, are combined in a logical OR to increase the overall efficiency. The triggers with the lower $\pT$ thresholds include isolation requirements on the lepton candidate, resulting in inefficiencies at high $\pT$ that are recovered by the triggers with higher $\pT$ thresholds.

\subsection{Preselection}
\label{sec:preselection}

Events are required to have exactly two leptons (electrons, muons, or both) with opposite charge, satisfying the criteria defined in Section~\ref{ch:object_definition}. Since single-lepton triggers are used, at least one of the two reconstructed leptons is required to have a $\pT>27$~\GeV and match a lepton with the same flavour reconstructed by the trigger algorithm within $\Delta R$ of 0.15. The chosen \pT threshold ensures a fully efficient trigger for the whole Run 2 period. In the $ee$ and $\mu\mu$ channels, the dilepton invariant mass must be above 15~\GeV and outside the $Z$ boson mass window 83--99~\GeV. Further suppression of the background is achieved by requiring at least three jets (either large-$R$ or small-$R$) with at least one of which being $b$-tagged using the DL1r 85\% WP\@. The fraction of signal events in the preselection region is negligible for all masses.

\subsection{Signal and control regions}
\label{sec:signal_control_regions}

After preselection, the data sample is dominated by background from $\ttbar$ events. To take advantage of the high jet and $b$-object multiplicities of the $\tta$ signal process, events are classified into non-overlapping regions based on the total number of $B$-jets and $b$-jets. Some of the regions also require the presence of at least one loose (and not tight) $b$-jet (small-$R$ jets tagged with the DL1r 85\%~WP but failing to meet the 70\% WP). The name of every signal region (SR) or control region (CR) includes the number of $B$-jets followed by ``B'' and of $b$-jets followed by ``b''. The names of those regions requiring at least an extra loose $b$-jet indicate it in their name with ``+1bL''.
Due to the high $b$-jet multiplicity of the signal, only regions with at least three $b$-objects are considered as signal regions ($B$-jets count as two $b$-objects). To maximise the signal sensitivity, signal events are classified into two boosted regions (SR 1B1b+1bL and SR 1B2b) and two resolved regions (SR 0B3b and SR 0B4b). The loose $b$-tagged jet in one of the signal regions is required to suppress $\ttbar$+light events.
The complementarity between boosted and resolved regions is illustrated in Figure~\ref{fig:invmass_SRs}, which shows the invariant mass of the $B$-jet in SR 1B2b and the invariant mass of the pair of $b$-jets with the largest $p_{T}$ in SR 0B4b for different values of the pseudoscalar mass. Regions containing one $B$-jet are particularly relevant in the boosted regime ($m_a < 30$~\GeV). Regions with no $B$-jets and a high $b$-jet multiplicity are more powerful in the resolved regime ($m_a > 30$~\GeV).

In addition to the four signal regions described above, a control region is included in the fit to improve the $\ttbar$+$\geq$1$c$ normalisation  (Section~\ref{ch:statistical_treatment}).
The control region (CR 0B2b+1bL), orthogonal to all four signal regions, is composed of events with no $B$-jets and exactly two $b$-jets, as well as at least one additional loose (and non-tight) $b$-jet. Similarly to the SR, the loose $b$-tagged jet is required to suppress $\ttbar$+light events that would otherwise dominate the control region. Finally, events entering the CR are required to have a sum of the pseudo-continuous $b$-tagging scores between 12 and 15. The number of $b$-jets in the 0B3b, 1B1b+1bL and 0B2b+1bL regions is exclusive, while it is inclusive in the 0B4b and 1B2b regions.
Table~\ref{tab:initial_SRs} summarises the selections for each region, which are applied
in addition to the previously mentioned preselection requirements.

\begin{figure}[tb]
\centering
\subfloat[]{\includegraphics[width=0.50\textwidth]{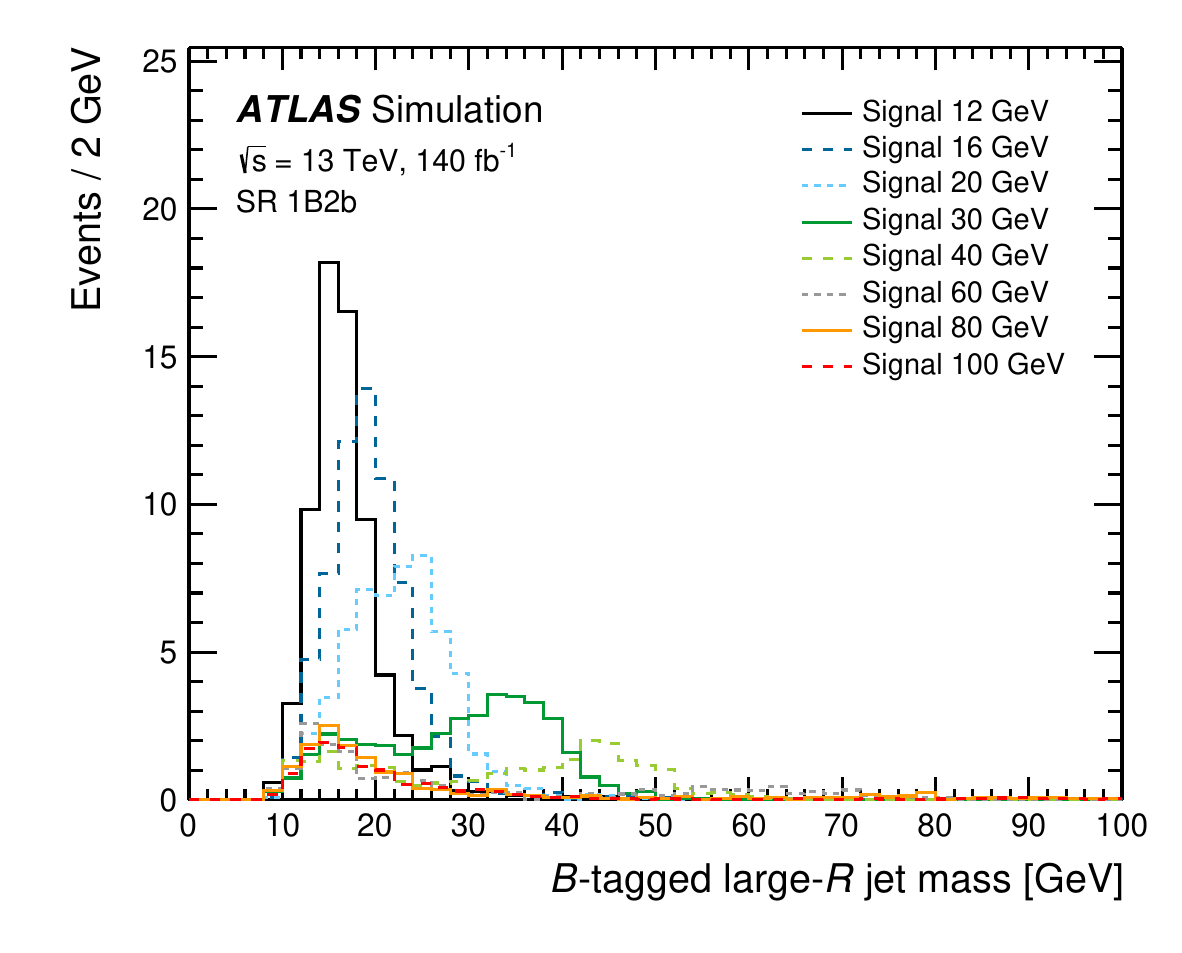}}
\subfloat[]{\includegraphics[width=0.50\textwidth]{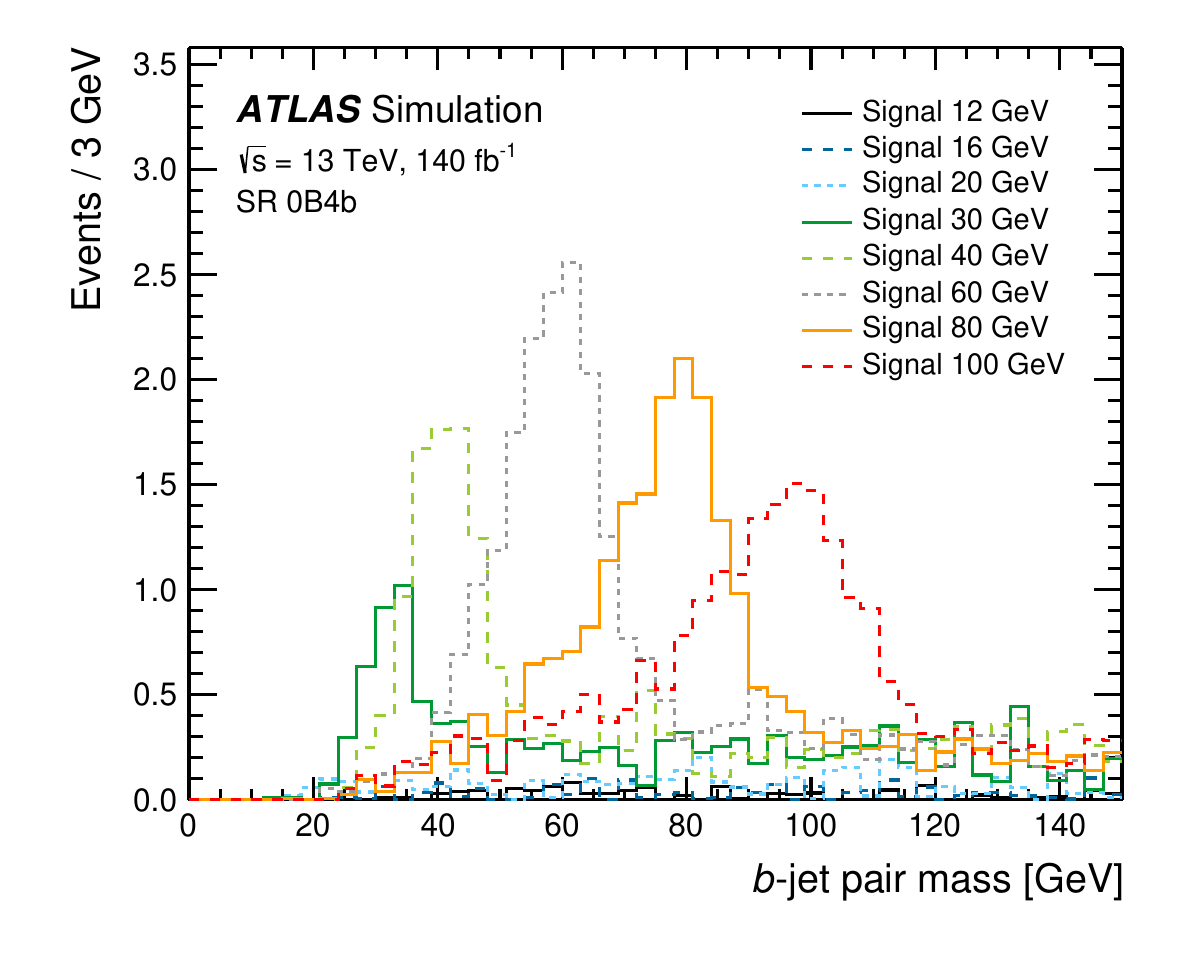}}

\caption{Invariant mass of the (a) $B$-jet in SR 1B2b and of the (b) pair of $b$-jets with the largest $p_\text{T}$ in SR 0B4b for various values of $m_{a}$. The distributions are normalised according to the predicted $m_{a}$-dependent theoretical cross sections with a Yukawa coupling of the $a$-boson to the top quark of 0.5.}
\label{fig:invmass_SRs}
\end{figure}

\begin{table}[h]
\caption{Overview of the jet multiplicities considered per region in the fit.}

\centering
\begin{tabular}{l|c|c|c|c|c}

Region & Large-$R$ jets & Small-$R$ jets & $B$-jets & $b$-jets & Loose $b$-jets \\ \hline
\hline
SR 0B4b & $\geq0$ & $\geq4$ & $=0$ & $\geq4$ & -\\
SR 0B3b & $\geq0$ & $\geq3$ & $=0$ & $=3$ & -\\
SR 1B2b & $\geq1$ & $\geq2$ & $=1$ & $\geq2$ & -\\
SR 1B1b+1bL & $\geq1$ & $\geq2$ & $=1$ & $=1$ & $\geq1$\\
CR 0B2b+1bL & $\geq0$ & $\geq3$ & $=0$ & $=2$ & $\geq1$\\

\multicolumn{2}{c}{}\\
\end{tabular}
\label{tab:initial_SRs}
\end{table}


\section{Background estimate}
\label{ch:reweighting}

Data-driven corrections are derived for the \ttbar Monte Carlo simulation, the main background process in this search. These corrections are derived to improve the description of the rates of \ttbar plus heavy flavour jets and the transverse momenta of lepton and jets observed in data, using a method similar to the one developed for other ATLAS searches~\cite{HDBS-2018-51, TOPQ-2020-10, HDBS-2020-16}. The corrections are derived in very inclusive control regions where the contamination of signal is predicted to be below 1\% for all considered pseudoscalar mass hypotheses.
The region of choice satisfies the preselection requirements described in Section~\ref{sec:preselection}, with an extra requirement of at least two $b$-jets and no $B$-jets. Additionally, to suppress the $Z$+jets contribution, only the different lepton flavour ($e\mu$) region is considered.

The first correction targets the production rate of heavy-flavour jets: $c$-jets and $b$-jets. It was observed in previous ATLAS and CMS analyses~\cite{CMS-TOP-11-007,CMS-TOP-18-002,TOPQ-2017-12} that the rate of $\ttbar$+HF events is underestimated in MC simulation. Due to the high $b$-object multiplicity of the $\tta$ signal, these HF events represent a large fraction of the $\ttbar$+jets background in the signal regions, and therefore the MC simulation must be corrected. To have a more accurate flavour composition, an event reweighting procedure is applied based on the \texttt{sumPCBTag} distribution.  Figure~\ref{fig:Correction_sumPCBTag}(a) shows how the $\ttbar$+light production is dominant at low values of \texttt{sumPCBTag}, while the $\ttbar$+HF production populates the tail of the distribution. The correction procedure consists in deriving three normalisation factors: one for each component, using a likelihood fit to the \texttt{sumPCBTag} MC distributions compared with data. In this fit, DL1r $b$-tagging systematic uncertainties (detailed in Section~\ref{ch:systematic_uncertainties}) are included. Figure~\ref{fig:Correction_sumPCBTag}(b) shows the improved agreement after the fit. The normalisation factors and their corresponding uncertainties are shown in Table~\ref{tab:Correction1_ttbarWt_NormFactors}.

This procedure also corrects the distribution of the number of jets (inclusive of all jet types) per event, $N_\text{jets}$, which was also observed to be mismodelled in the simulation. Figure~\ref{fig:Correction_sumPCBTag_Njets} shows the corresponding distribution, before and after applying the normalisation factors from Table~\ref{tab:Correction1_ttbarWt_NormFactors}. In the following, in all Figures and Tables, the category ``Other'' includes the following minor background processes: $Z$+jets, $\ttbar W$, $tq$, $tZ$, $tWZ$, $WW$, $ZZ$, $WZ$ and $W$+jets.

\begin{figure}[tb]
\centering
\subfloat[]{\includegraphics[width=0.5\textwidth]{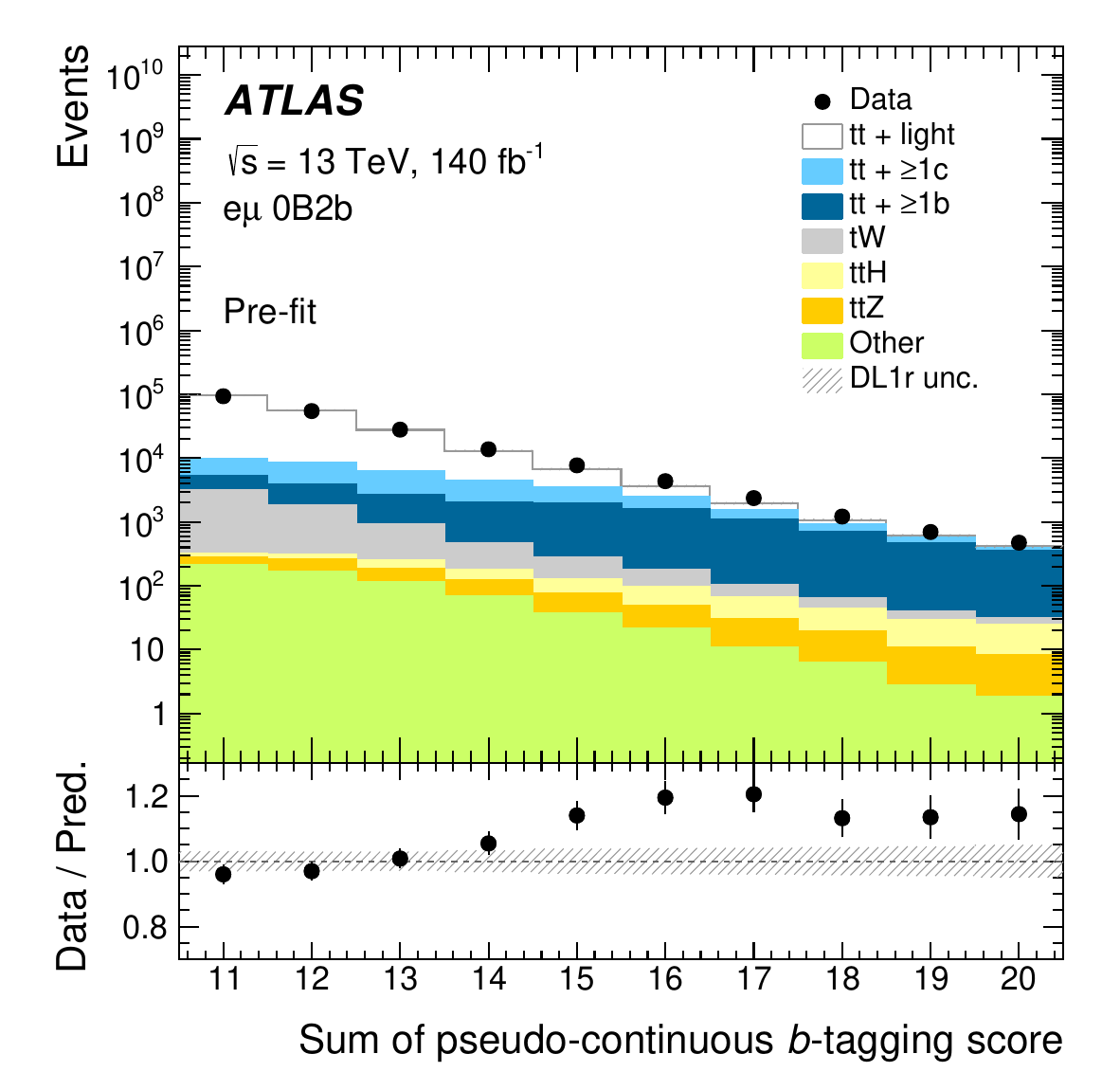}}
\subfloat[]{\includegraphics[width=0.5\textwidth]{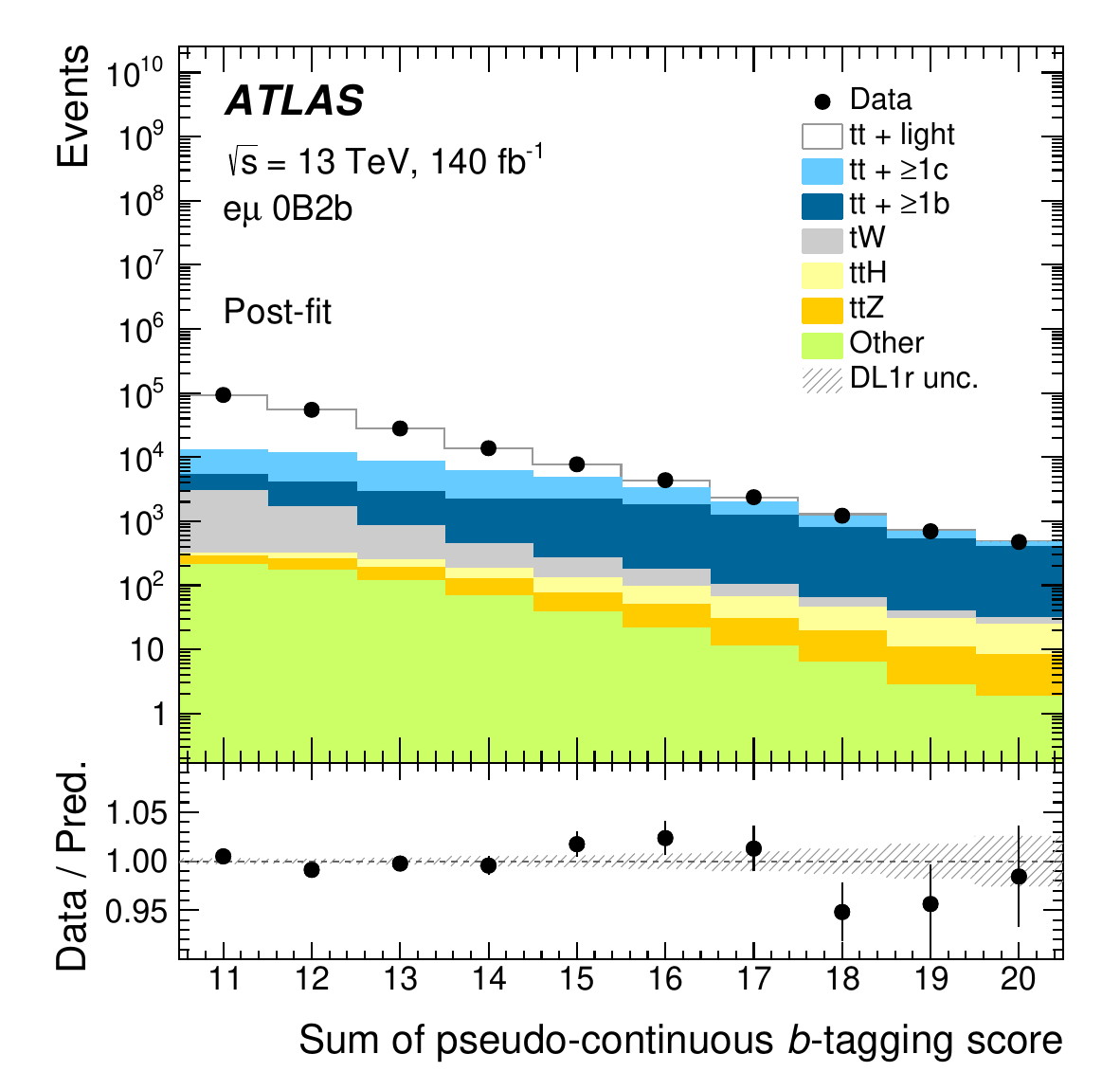}}
\caption{Comparison of the data versus MC distribution of the sum of the pseudo-continuous $b$-tagging score of all the jets per event (a) before and (b) after applying the normalisation factors extracted from the heavy-flavour correction fit.}
\label{fig:Correction_sumPCBTag}
\end{figure}

\begin{table}[tb]
\caption{Normalisation factors for the three $t\bar{t}$+jets HF categories resulting from the likelihood fit performed using the \texttt{sumPCBTag} distribution.}
\centering
\begin{tabular}{l|c}

HF category & Norm. factor \\ \hline
\hline
$t\bar{t}$+light, $tW$ & 0.91 $\pm$ 0.03 \\
$t\bar{t}$+$\geq$1$c$ & 1.58 $\pm$ 0.14 \\
$t\bar{t}$+$\geq$1$b$ & 1.13 $\pm$ 0.07 \\ \multicolumn{2}{c}{}\\
\end{tabular}

\label{tab:Correction1_ttbarWt_NormFactors}
\end{table}

\begin{figure}[tb]
\centering
\subfloat[]{\includegraphics[width=0.5\textwidth]{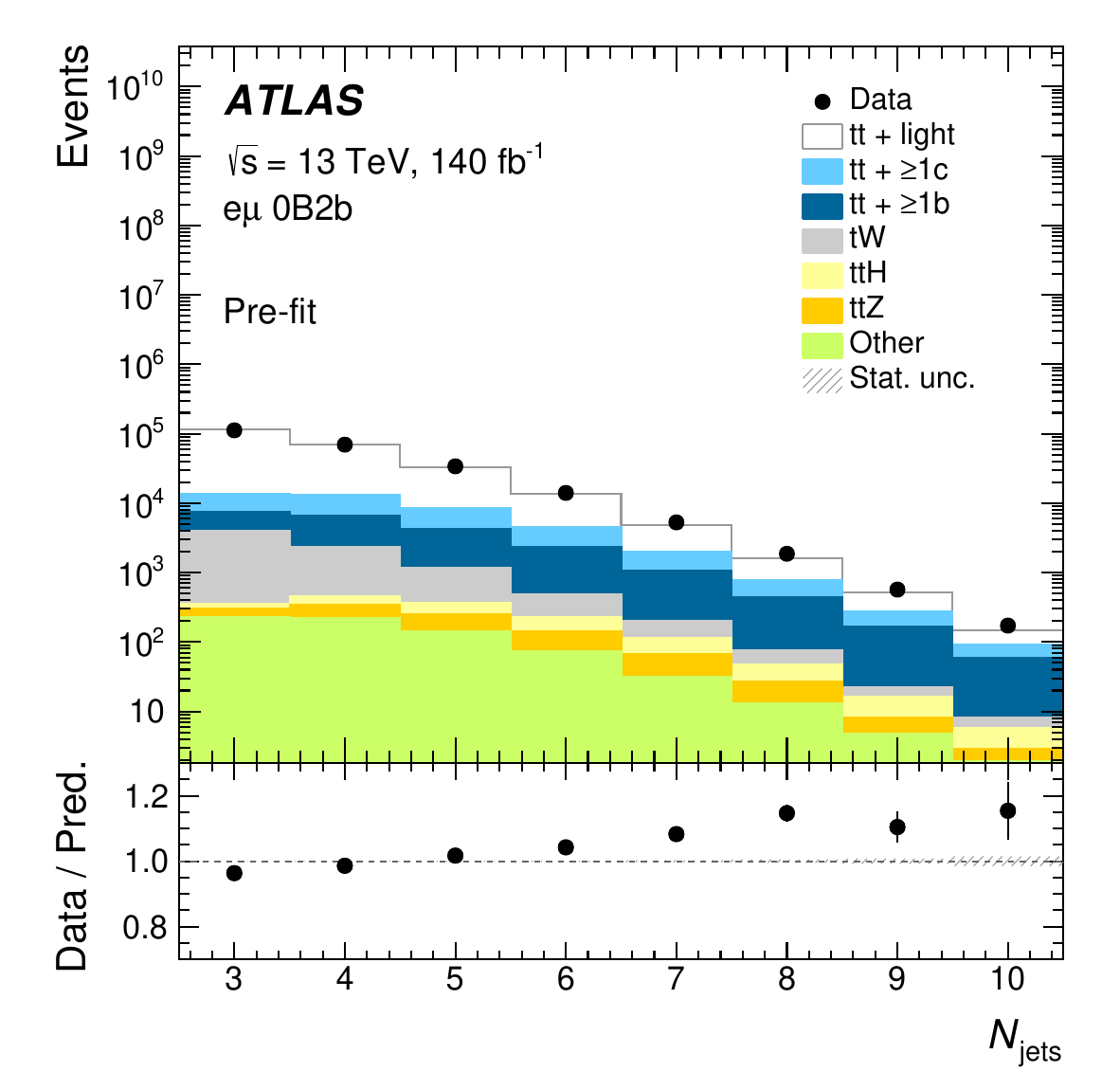}}
\subfloat[]{\includegraphics[width=0.5\textwidth]{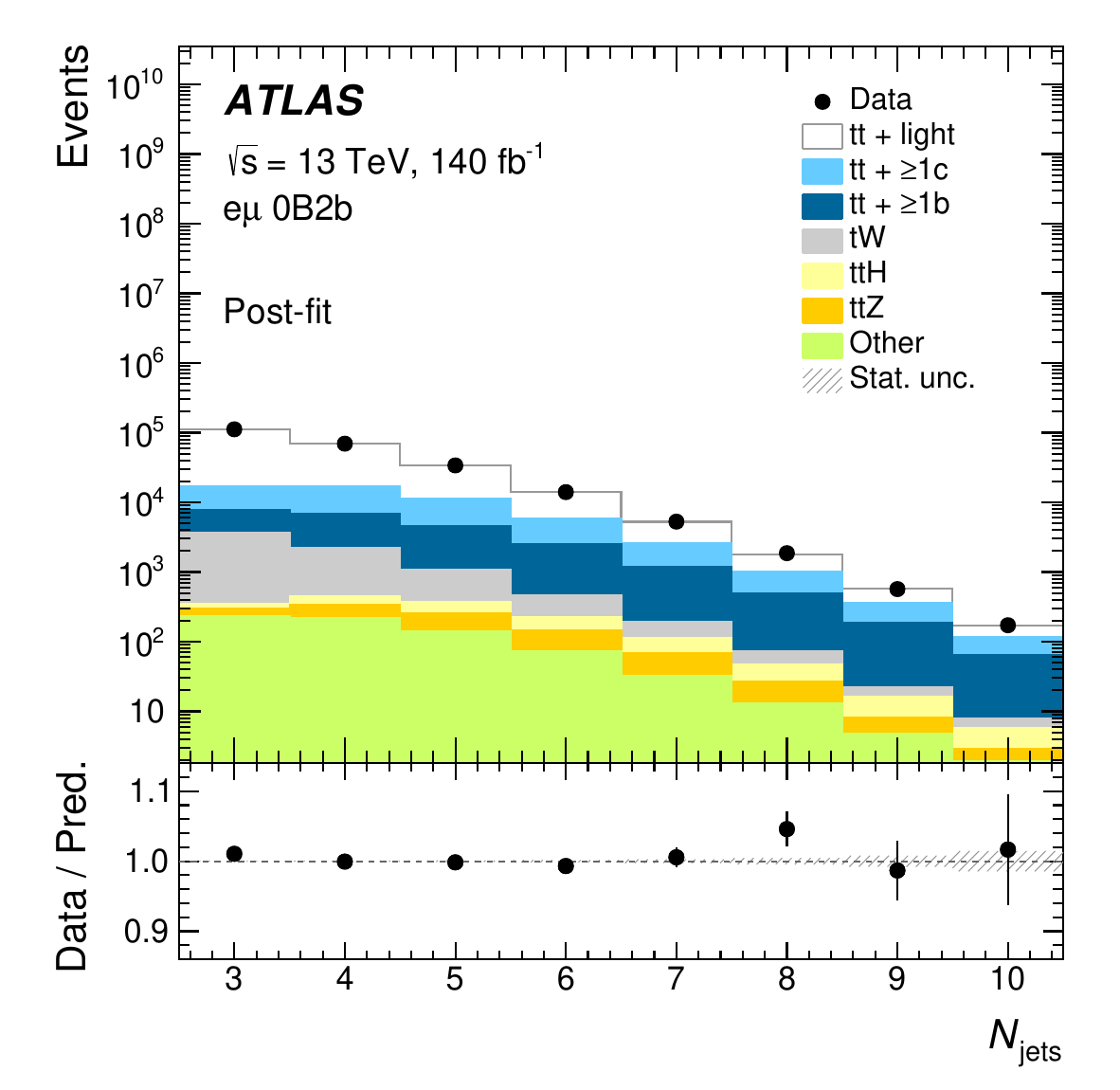}}
\caption{Comparison of the data versus MC $N_\text{jets}$ distribution (a) before and (b) after applying the $t\bar{t}$+jets normalisation factors extracted from the heavy flavour correction fit. Only statistical uncertainties are shown.}
\label{fig:Correction_sumPCBTag_Njets}
\end{figure}

The second correction targets the transverse momenta of the jets and leptons originating from the decay of top quark/antiquark, a quantity that was also observed to be mismodelled by current \ttbar 5FS MC generators. The disagreement between data and MC persists even after applying the $\ttbar$+HF correction. To improve the agreement between data and MC for these variables, a kinematic reweighting factor for the $\ttbar$+$\geq$1$c$, $\ttbar$+light, and $tW$ components is derived from the data/MC ratio, after subtracting other background components from the data. These mismodellings are assumed to be independent of the flavour of the extra radiation, and applied equally to $\ttbar$+light, $\ttbar$+$\geq$1$c$, and $tW$.

The event hardness or $H_\text{T}$, which is defined as the scalar sum of the $p_\text{T}$ of all the jets and leptons in the event, is largely correlated with the total number of jets in the event, as every additional jet in the event shifts $H_\text{T}$ to larger values.
Performing the kinematic reweighting directly with $H_\text{T}$ would therefore spoil the data/MC agreement achieved after the HF correction to the number of jets distribution shown in Figure~\ref{fig:Correction_sumPCBTag_Njets}.
To reduce the $N_\text{jets}$ dependency, a new variable, $H_{\text{T}}^\text{red}(n)$, is defined:
\begin{equation*}
H_\text{T}^\text{red}(n) = H_\text{T} - (n-3)\Delta H_\text{T}(n),
\end{equation*}
where $n$ is the number of jets (small-$R$ and large-$R$ jets, with a minimum of three jets) and $\Delta H_\text{T}(n)$ is the average offset in $H_\text{T}$ caused by the addition of each extra jet to the event.
Correction factors are derived over a binned $H_\text{T}^\text{red}$ distribution, and the results are fitted using a continuous hyperbolic function which is later used for the MC reweighting.
Figure~\ref{fig:HT_ETmiss_HT_correction} shows the $H_\text{T}$ distribution before and after applying the correction. Similar improvements are observed for individual leptons and jets. No significant changes in the number of jets distribution are observed after the kinematic corrections are applied.
Following the same procedure, dedicated kinematic reweighting corrections are derived for the alternative $\ttbar$+light, $\ttbar$+$\geq$1$c$, and $tW$ samples employed in the  evaluation of systematic uncertainties described in Section~\ref{ch:systematic_uncertainties}. Residual differences between data and the MC are taken into account in the analysis fit by including free-floating individual normalisations for the various $\ttbar$+jets background contributions.

\begin{figure}[htb]
\centering
\subfloat[]{\includegraphics[width=0.5\textwidth]{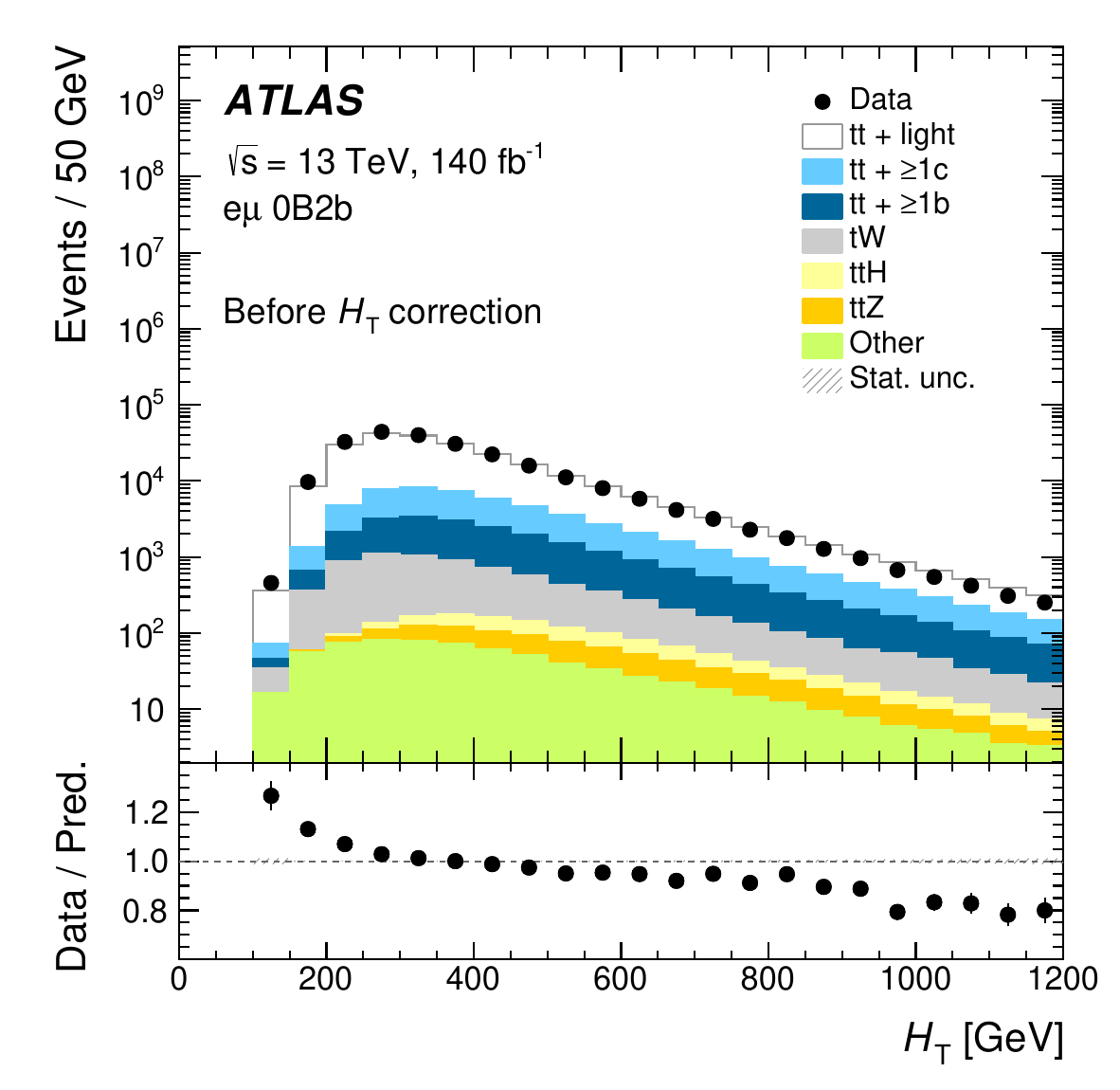}
}
\subfloat[]{\includegraphics[width=0.5\textwidth]{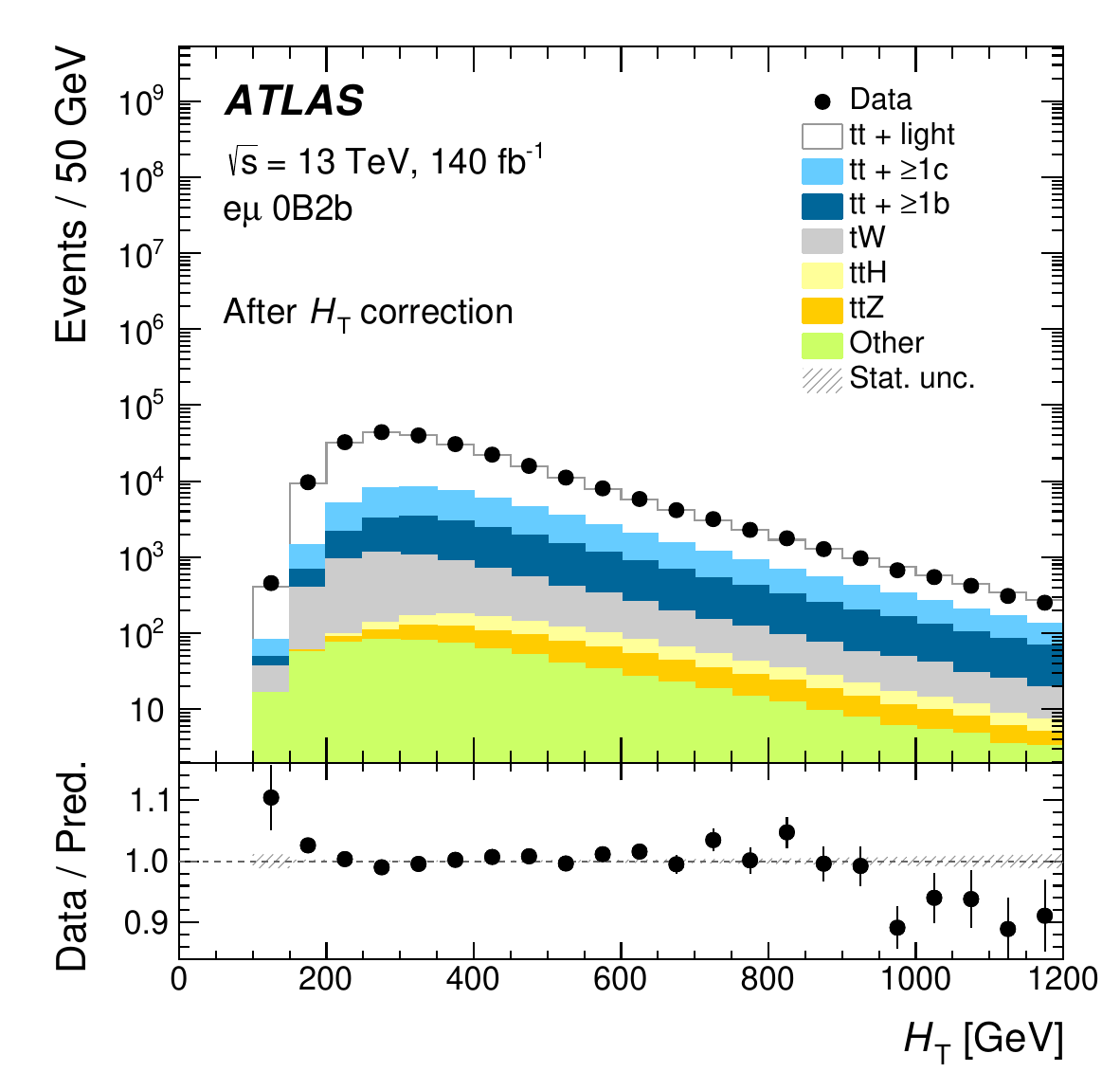}}

\caption{Comparison between data and MC of the $H_\text{T}$ distribution (a) before and (b) after correcting it using $H_\text{T}^\text{red}$. Only statistical uncertainties are shown.}
\label{fig:HT_ETmiss_HT_correction}
\end{figure}


\section{Analysis strategy}
\label{ch:analysis_strategy}

The analysis uses various machine learning (ML) algorithms to improve the sensitivity to the target signal. First, two different boosted decision trees (BDTs) are trained to identify the jets originating from the decay of the top quark and antiquark and the jets originating from the pseudoscalar decay.
Second, a mass-parameterised NN is trained for signal/background discrimination in each of the SRs described in Section~\ref{sec:signal_control_regions}.
The final fit uses the NN output score distribution in each of the four SRs and the \texttt{sumPCBTag} distribution in the CR.
In addition to the signal strength $\mu$, three normalisation factors, corresponding to the main background contributions, are left to freely float in the fit. Figure~\ref{fig:AnalysisDiagram} shows a diagram summarising the ML approach followed in the analysis as well as the CR and SRs used in the final fit. Further details on each step are given in the following.

\begin{figure}[tb]
\centering
\subfloat{\includegraphics[width=0.9\textwidth]{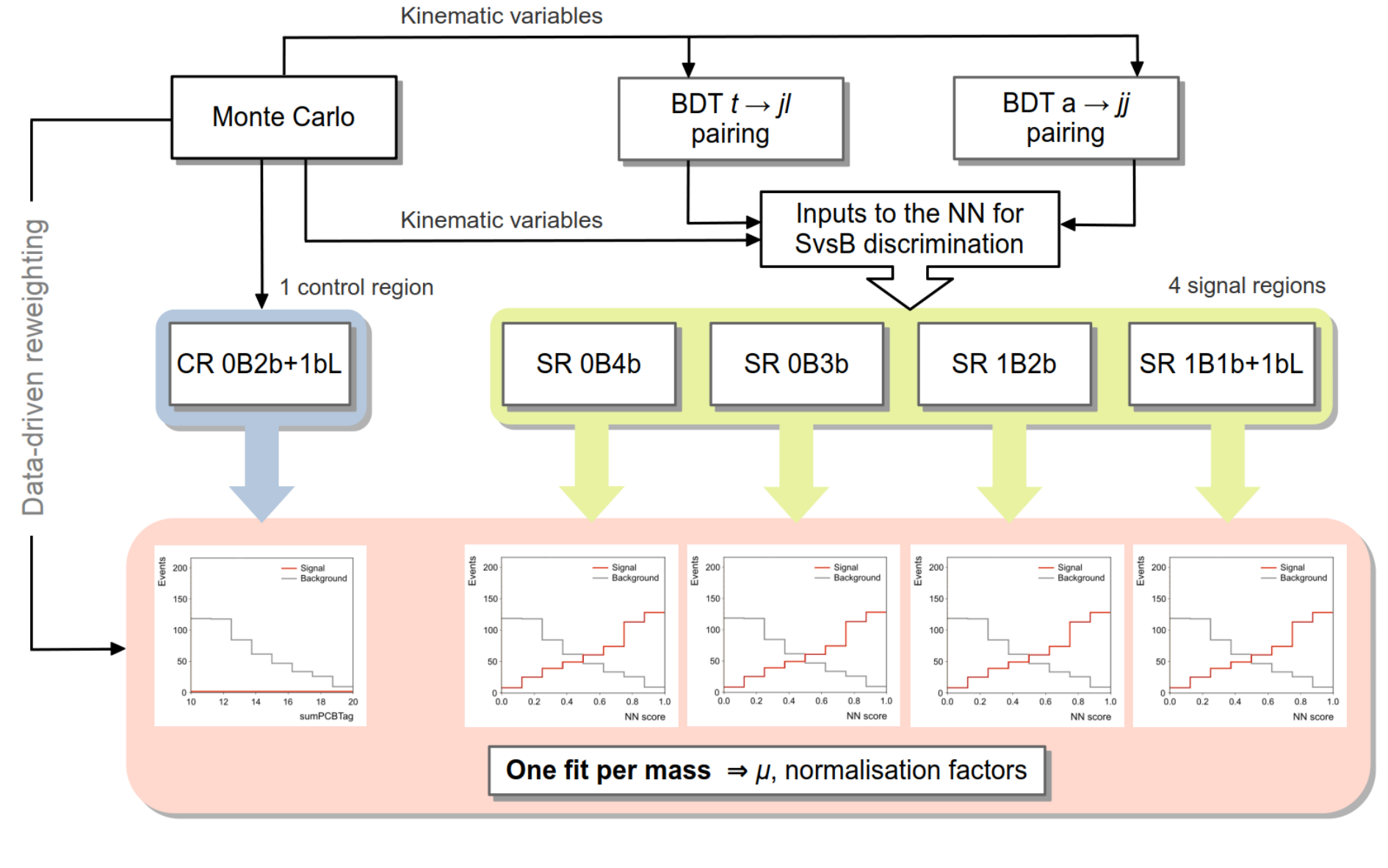}}
\caption{Diagram of the analysis strategy, illustrating the data-driven corrections, jet/lepton and dijet pairings by two BDTs and the NN for signal versus background discrimination. The CR and SRs are used in the final fits to extract the signal strength ($\mu$) as well as normalisation factors for the main backgrounds.}
\label{fig:AnalysisDiagram}
\end{figure}

\subsection{Event reconstruction BDTs}
\label{ch:BDT}

Two different BDTs are trained to do partial event reconstruction. One targets the identification of the lepton-jet pairs associated with the top quark/antiquark decays, while the other identifies the pair of jets from the pseudoscalar decay.
The two BDTs use the five leading small-$R$ jets, together with the two charged leptons in the case of the top quark/antiquark BDT, and in each case select the pair of jets or the lepton/jet pair most likely to correspond to the pseudoscalar or the top quark/antiquark, respectively. Both BDTs were designed to reconstruct resolved topologies, thus they do not use large-$R$ track-jets. Also, no attempt to reconstruct the two neutrinos from the top quark/antiquark decay is made.
The two BDTs are trained using all signal samples inclusively, such that they are generic and valid for all considered masses. During the BDT training process, target labels for each jet are assigned based on a one-to-one matching between reconstructed jets at the detector level and parton-level $b$-quarks/leptons. Consequently, a reconstructed jet/lepton is assigned as originating from a (anti-)top quark, or pseudoscalar decay candidate based on the aforementioned generator information.

The BDT targeting the top quark/antiquark decay attempts to pair each lepton with its corresponding $b$-jet, considering each lepton in turn. For this, the BDT receives as input several kinematic variables that depend on the tag lepton/jet pair ($jl$-pair), such as its invariant mass or transverse momentum, or the separation angle between the lepton and the jet. It also uses information about the lepton and jet candidates themselves, such as their pseudorapidity and transverse momenta, or the jet index indicating in decreasing order the hardness of the jets. In addition, the BDT uses information from the auxiliary $jl$-pair built with the lepton that is not being evaluated, together with information from the top/antitop system formed by the tag and auxiliary $jl$-pairs, and variables that refer to the full event.
In a similar way, the BDT targeting the pseudoscalar decay receives various kinematic variables connected to the pair of two jets, $j\!j$-pair, as its mass or transverse momentum, together with information about the jets themselves or about the overall event. The full list of variables used by each BDT is shown in Table~\ref{tab:BDT_variables}.

\begin{table}[tb]
\caption{Input variables used in the (a) top quark/antiquark reconstruction BDT and (b) pseudoscalar reconstruction BDT.}
\centering
\begin{tabular}{cc}
\begin{minipage}{.55\linewidth}
\begin{tabular}{r|l}
\multicolumn{2}{c}{(a)}\\
\multicolumn{2}{c}{Top quark/antiquark reconstruction BDT}\\

Object & Variables  \\ \hline
\hline
Full event & $N_{\text{jets}}$, $N_{b\text{-jets}}$\\
Lepton (tag, aux.) &  $p_\text{T}$, $\eta$  \\
Jet (tag, aux.) &  $p_\text{T}$, $\eta$, PC $b$-tag, jet index   \\
$jl$ pair (tag, aux.) & $m$, $p_\text{T}$, $\eta$,  $\Delta R$  \\
$\ttbar$ pair & $m$, $p_\text{T}$, $\eta$, $\Delta R$, $\Delta \phi$\\
$j\!j$ pair & $\Delta R$ \\
\end{tabular}
\end{minipage}%
\begin{minipage}{.54\linewidth}
\begin{tabular}{r|l}
\multicolumn{2}{c}{ }\\

\multicolumn{2}{c}{(b)}\\
\multicolumn{2}{c}{Pseudoscalar reconstruction BDT}\\

Object & Variables  \\ \hline
\hline
Full event & $N_{\text{jets}}$, $N_{b\text{-jets}}$, \texttt{sumPCBTag}\\
Jet (1st, 2nd) &  $p_\text{T}$, $\eta$, PC $b$-tag, jet index   \\
$j\!j$ pair & $m$, $p_\text{T}$, $\eta$, $E$, $\phi$, $\Delta R$  \\

\multicolumn{2}{c}{ }\\
\multicolumn{2}{c}{ }\\
\multicolumn{2}{c}{ }\\
\multicolumn{2}{c}{ }\\
\end{tabular}
\end{minipage}
\end{tabular}

\label{tab:BDT_variables}
\end{table}

For the training of both BDTs, generator information is used to define the targets ($b$-quarks and leptons from the decays of the top quark, top antiquark and pseudoscalar $a$) and to identify the correct permutations at detector level, which are used as signal (or target) during the training, while all wrong permutations are used as background. A mix of signal and \ttbar samples is used during the training of both BDTs.
In both cases, two sets of BDTs are trained using k-2 fold training to avoid biases. The BDT trained with odd events is applied to even events and vice versa.
The training is performed with the TMVA package of ROOT~\cite{TMVA}.

Following the training of both BDTs, they are applied to data and MC as follows. For the top quark/antiquark BDT, the lepton/jet permutation with the highest BDT score is identified for each lepton as the most likely $jl$-pair and the selected jet is assigned to the top quark or antitop quark decay depending on the lepton charge. If both leptons are initially assigned to the same jet, only the one with the highest BDT score keeps the assignment, while the other lepton is reassigned to the second most likely jet in terms of BDT score. In a similar way, for the pseudoscalar BDT, the permutation of two jets with the highest BDT score is selected and the two corresponding jets are assigned to the pseudoscalar decay. The selected $jl$- and $j\!j$-pairs are used to define related variables, such as the top-quark or pseudoscalar reconstructed invariant mass or separation angles, that are later used as input by the signal-versus-background discrimination neural networks. Figure~\ref{fig:BDT_InvMasses} shows the prefit data/MC comparison of the reconstructed mass of the $lb$-pair selected by the top-quark BDT and of the $j\!j$-pair selected by the pseudoscalar BDT in SR 0B3b, the signal region with the largest statistics.

\begin{figure}[tb]
\centering
\subfloat[]{
\includegraphics[width = 0.48\textwidth]{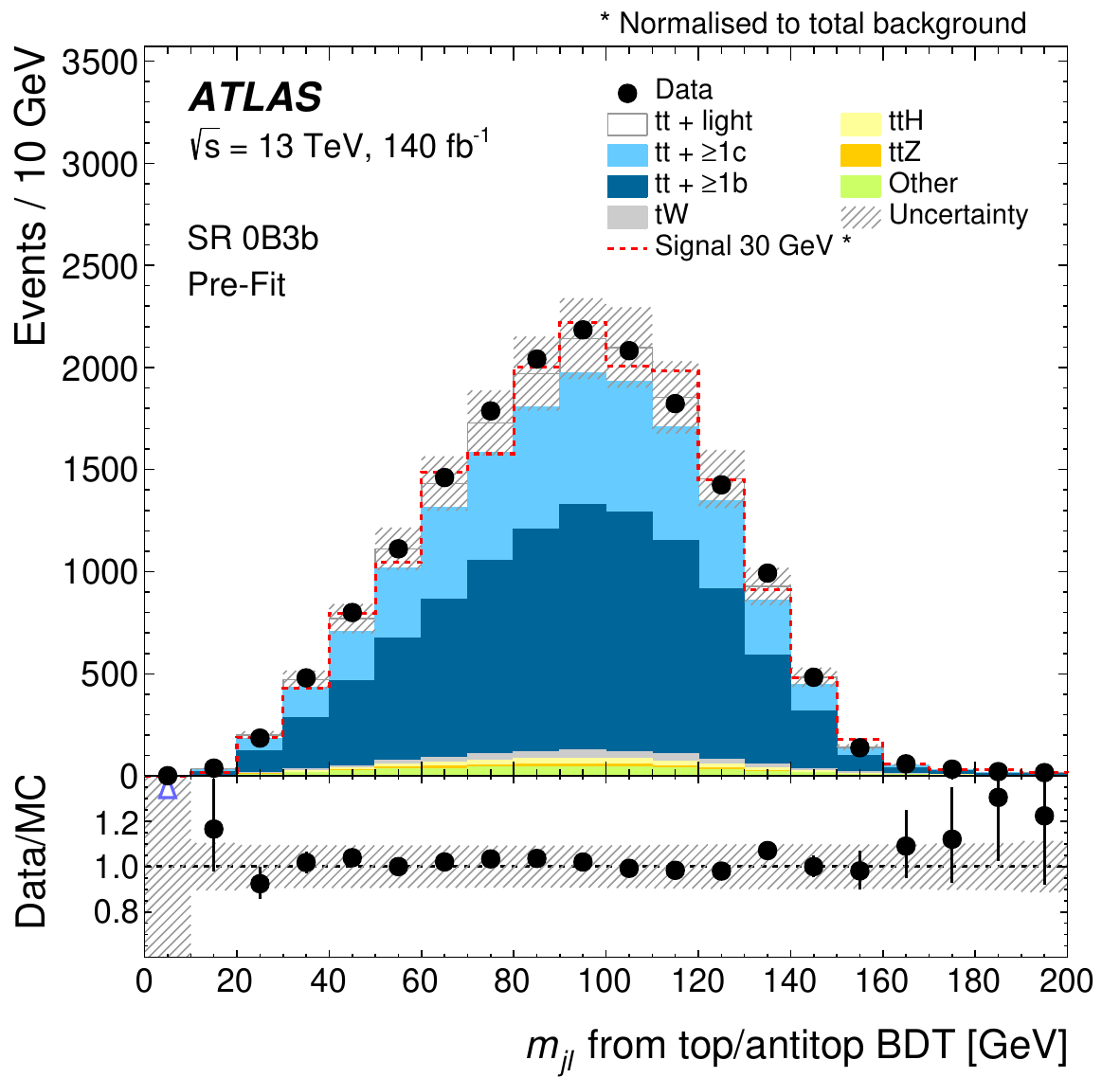}}
\subfloat[]{
\includegraphics[width = 0.48\textwidth]{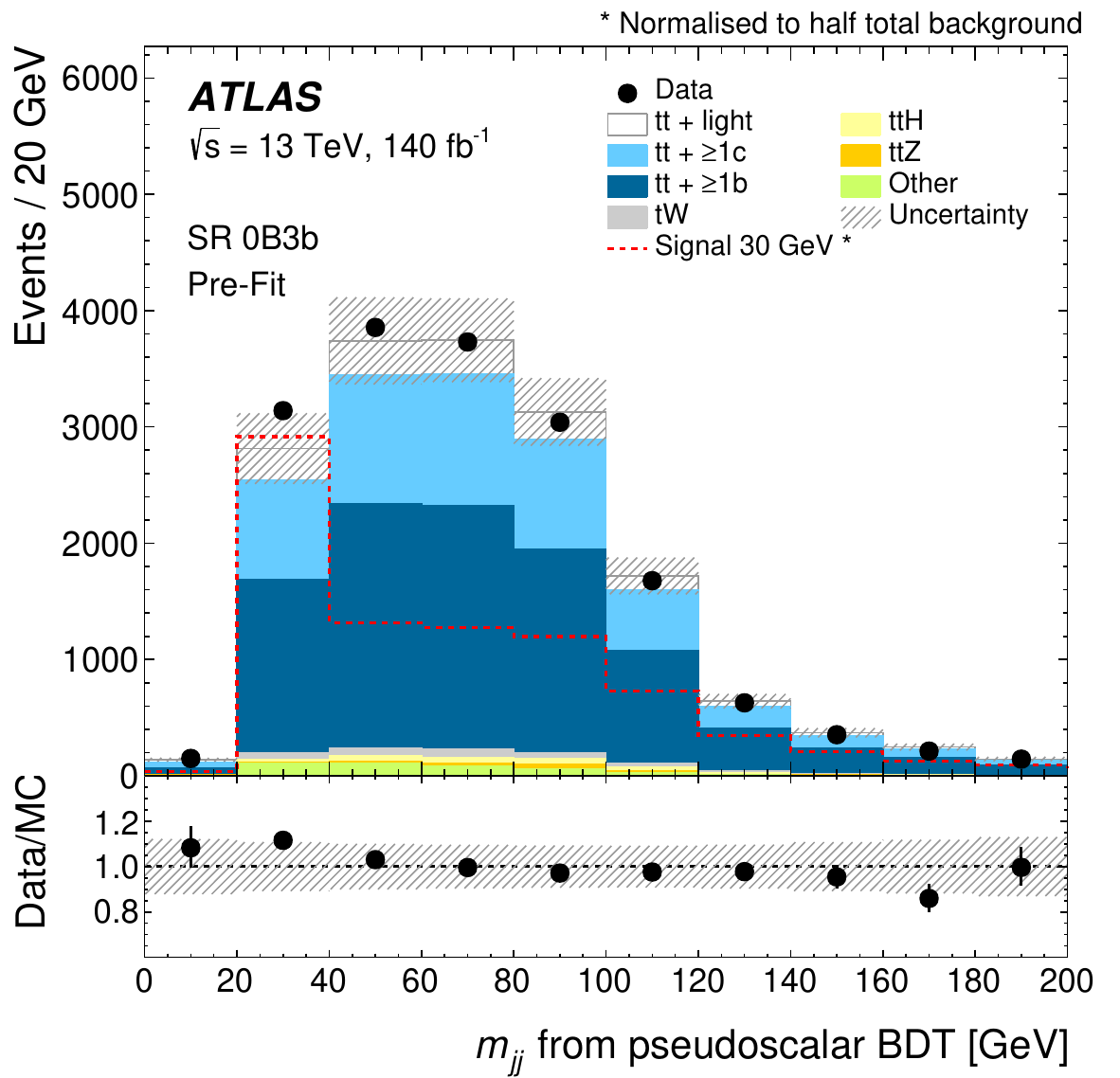}}
\caption{Distribution (a) of the antitop-quark $m_{jl}$ and (b) pseudoscalar $m_{jj}$ selected by the top quark and pseudoscalar BDTs, respectively in SR 0B3b before the fit. The dashed line shows the distribution of the 30~\GeV signal normalised to the total or half the total number of events in the region. The band displays the total pre-fit uncertainty.}
\label{fig:BDT_InvMasses}
\end{figure}

\subsection{Signal-versus-background discriminating neural networks}
\label{ch:NN}

As described in Section~\ref{sec:signal_control_regions}, events are divided into four signal regions according to their $B$- and $b$-jet multiplicity to better separate signal from background. The four signal regions used in the final fit are SR 0B4b, SR 0B3b, SR 1B2b and SR 1B1b+1bL. Independent NNs are trained individually per region to separate signal from background.

To make better use of the MC samples in the training, five different trainings are performed independently per region, where 80\% of the events in the region constitute the training sample and the remaining 20\% are used as the validation sample.
An appropriate distribution of events in the various samples guarantees that no event is used both in the training and the evaluation of the NNs.

Each NN contains two hidden layers with twice as many nodes as the input layer, connected by Rectified Linear Unit (ReLU) activation functions.
The final layer is a single node, normalised by a sigmoid function. The dropout for every layer is set to 0.3. To avoid overtraining, early stopping is implemented when the validation loss function does not improve during the last four epochs. The training is done using PyTorch 1.13.1~\cite{Paszke:2019xhz}, and each of the NNs combines basic four-momentum information with high-level variables, such as invariant masses or angular distances, as well as relevant variables from the BDTs described in Section~\ref{ch:BDT}. The full list of input variables depends slightly on the region, given the slightly different signal topologies per region. Table~\ref{tab:NN_input_per_region_aux} shows the overall list of input variables used by the NNs. Some of the most important variables in the NNs are $H_{\mathrm{T}}^{\mathrm{jets}}$, the invariant mass of two small-$R$ jets or the mass and \pT of the large-$R$ jet, among others.

\begin{table}[tb]
\caption{List of input variables to the NNs. The distributions corresponding to both the pair with the maximum $p_\text{T}$ and minimum $\Delta R$ are included for $bb$ variables. Angular variables with one $b$ or one $B$ use the pair with the minimum $\Delta R$. The $m_{bbbb}$ and $m_{bbb}$ variables correspond to the combination with the maximum scalar sum of $p_\text{T}$.}
\centering
\begin{tabular}{r|l}
\multicolumn{2}{c}{Signal-versus-background discrimination NN}\\

Object & Variables \\ \hline \hline
Full event & $N_\text{jets}$, $H^\text{jets}_\text{T}$, $E_\text{T}^\text{miss}$ \\
BDT $t\rightarrow jl$ & Score, $p_\text{T}^{jl}$, $\Delta R_{jl}$, $\Delta\eta_{jl}$, $\Delta\phi_{jl}$, jet index \\
BDT $a\rightarrow jj$ & Score, $p_\text{T}^{jj}$, $\eta_{jj}$, $m_{jj}$, $\Delta R_{jj}$, $\Delta\eta_{jj}$, $\Delta\phi_{jj}$, jet index\\
Leptons & $\Delta R_{ll}$, $\Delta\eta_{ll}$, $\Delta\phi_{ll}$, $\Delta\phi_{E_\text{T}^\text{miss},l}$,  $\Delta R_{ll,bb}$, $\Delta R_{ll, B}$, $\Delta R_{ll,b}$\\

Large-$R$ jets & $p_\text{T}$, $\eta$, $m$, $\Delta R_{Bb}$, $\Delta\phi_{E_\text{T}^\text{miss},B}$ \\

\multirow{2}{6em}{\raggedleft Small-$R$ jets}
& $p_\text{T}^{bb}$, $m_{bb}$, $m_{bbb}$, $m_{bbbb}$, $\Delta R_{bb}$, $\Delta\eta_{bb}$, $\Delta\phi_{bb}$, $\Delta\phi_{E_\text{T}^\text{miss},b}$ \\
& $p_\text{T}$, $\eta$, PC $b$-tag \\

\multicolumn{2}{c}{}\\

\end{tabular}

\label{tab:NN_input_per_region_aux}
\end{table}

All NNs are mass-parameterised, meaning that they receive the mass hypothesis as input during the evaluation. For the training, background MC samples are randomly assigned a mass from the grid of generated signal samples, while appropriate mass values are assigned to the signal events. Once the NNs are trained, the data scores are evaluated for each value of $m_a$ considered in the analysis. In each SR, the resulting NN score is the distribution used in the profile likelihood fit, as discussed in the next Section. Figure~\ref{fig:sec_Asimov_tta_30_NNscore_prefit} shows the prefit distributions of the four NN scores corresponding to the 30~\GeV mass hypothesis.

\begin{figure}[tb]
\centering\subfloat[]{
\includegraphics[width=0.40\textwidth]{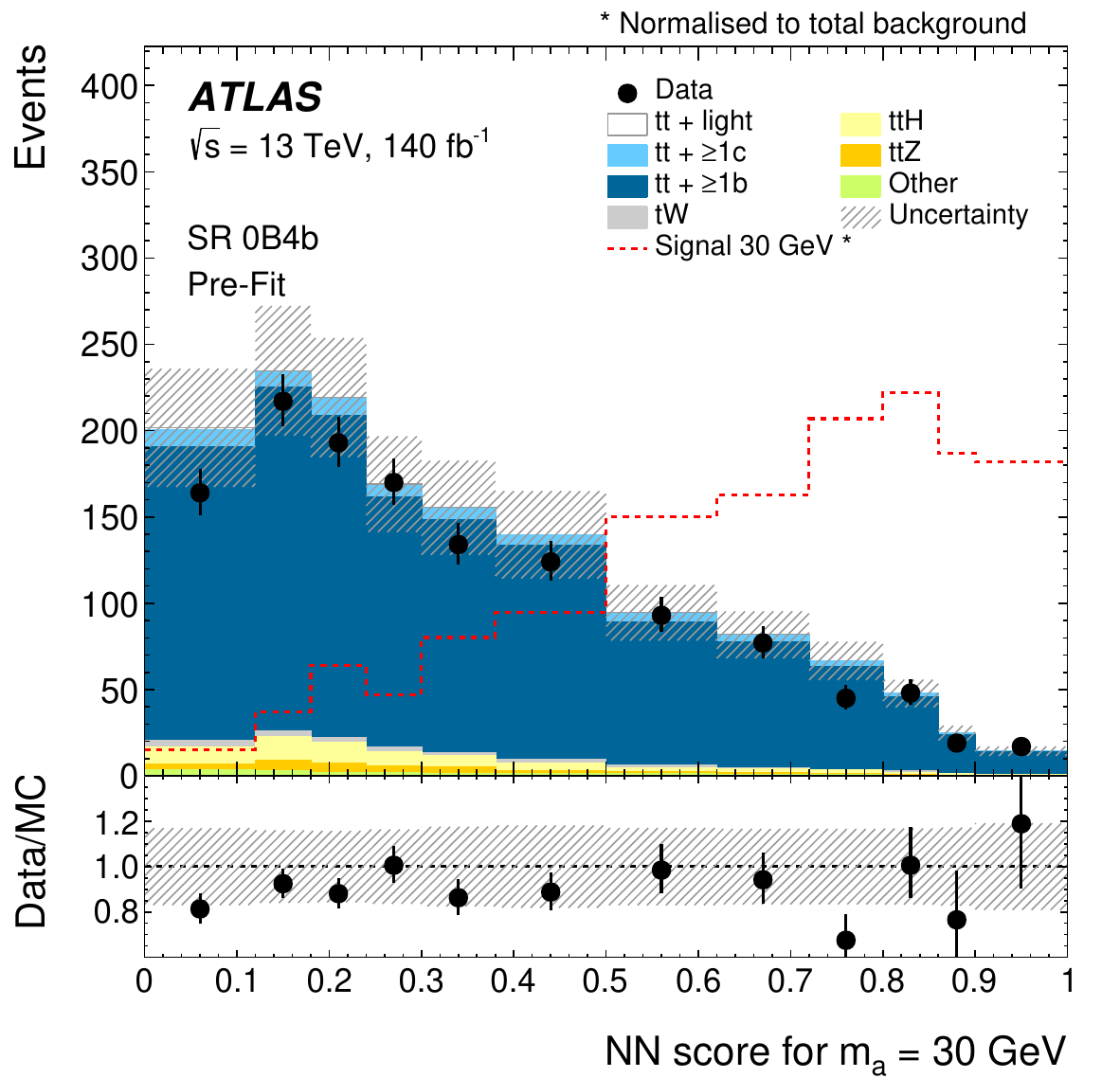}}
\subfloat[]{\includegraphics[width=0.40\textwidth]{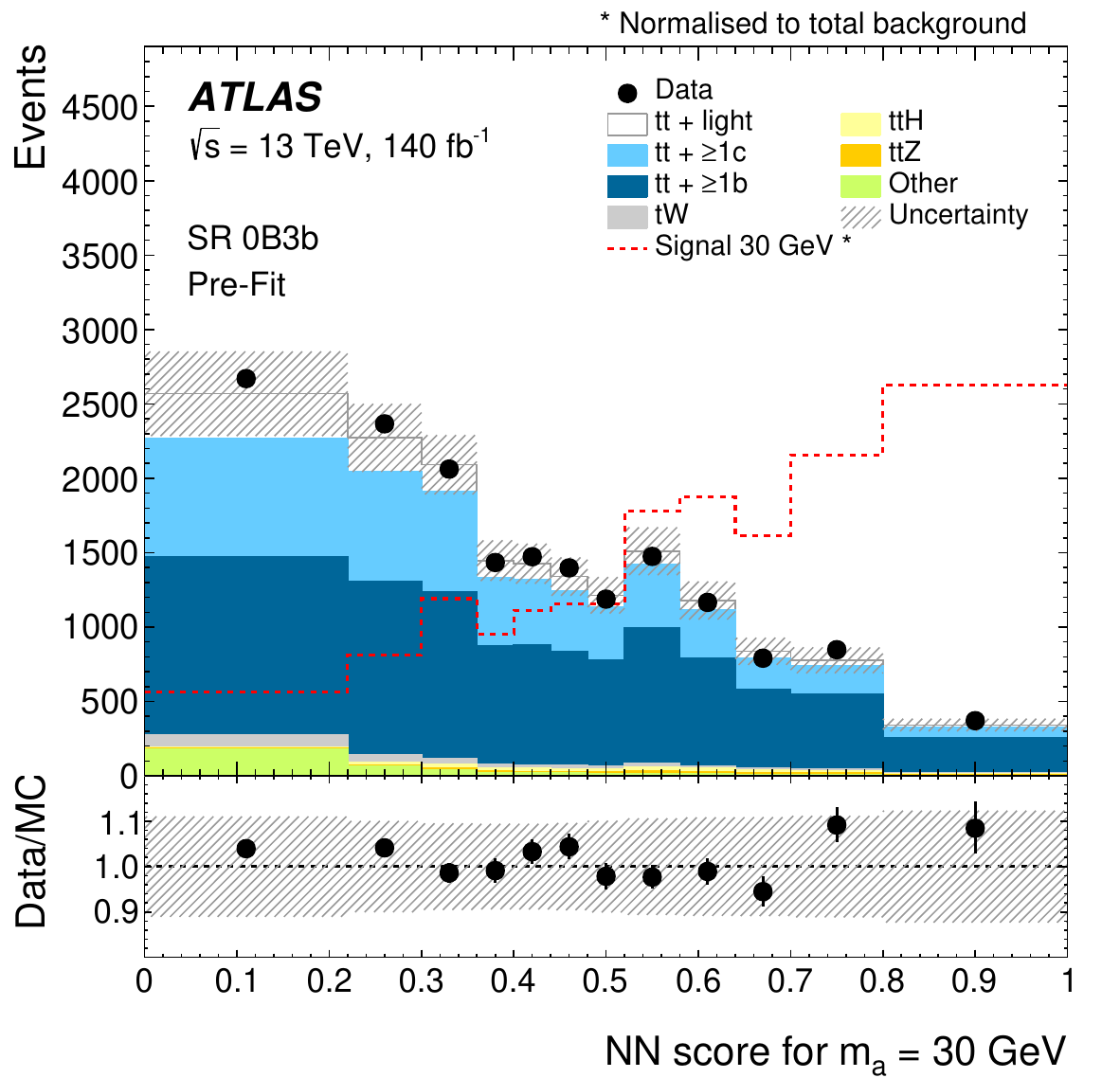}} \\
\subfloat[]{\includegraphics[width=0.40\textwidth]{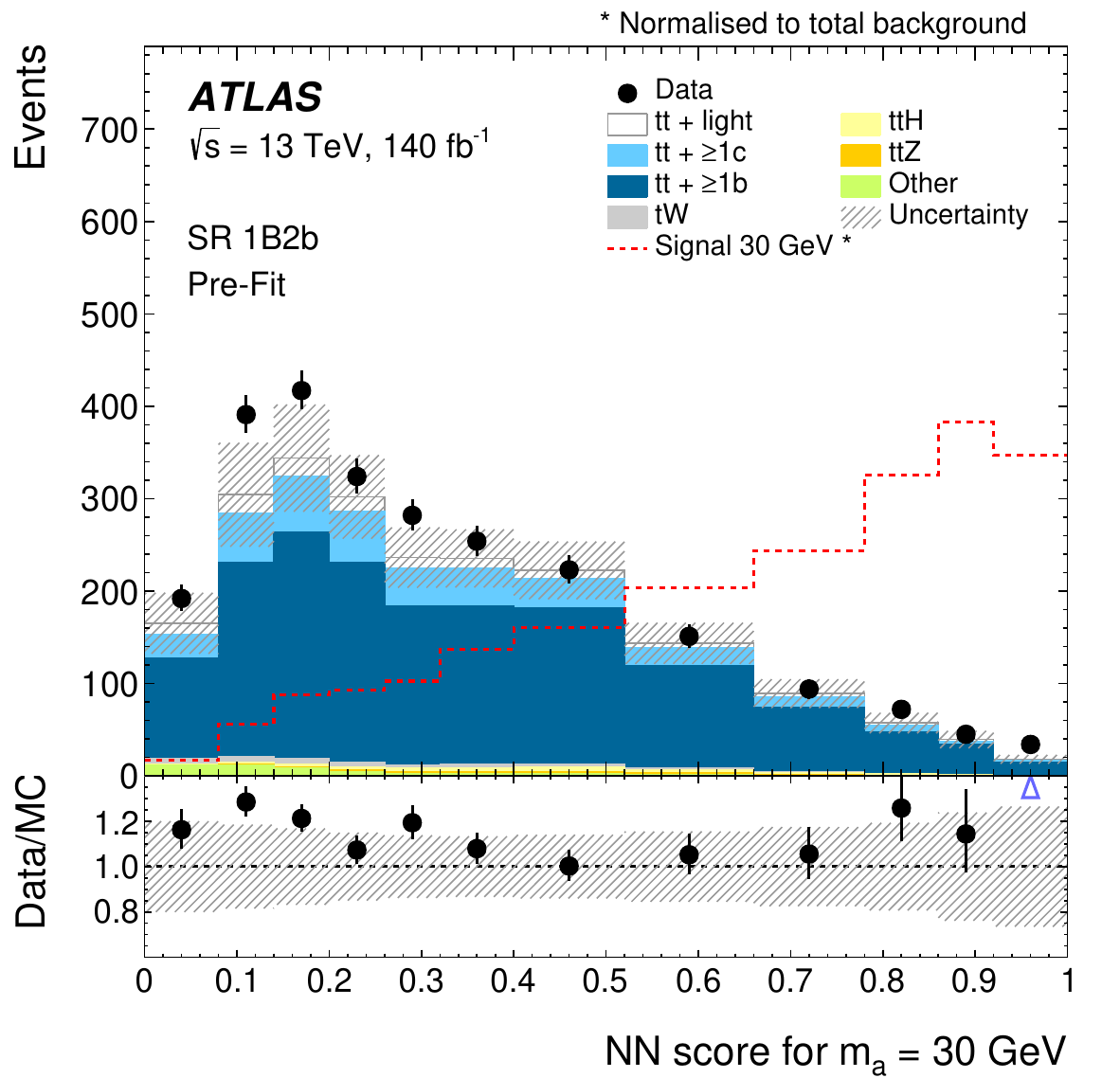}}
\subfloat[]{     \includegraphics[width=0.40\textwidth]{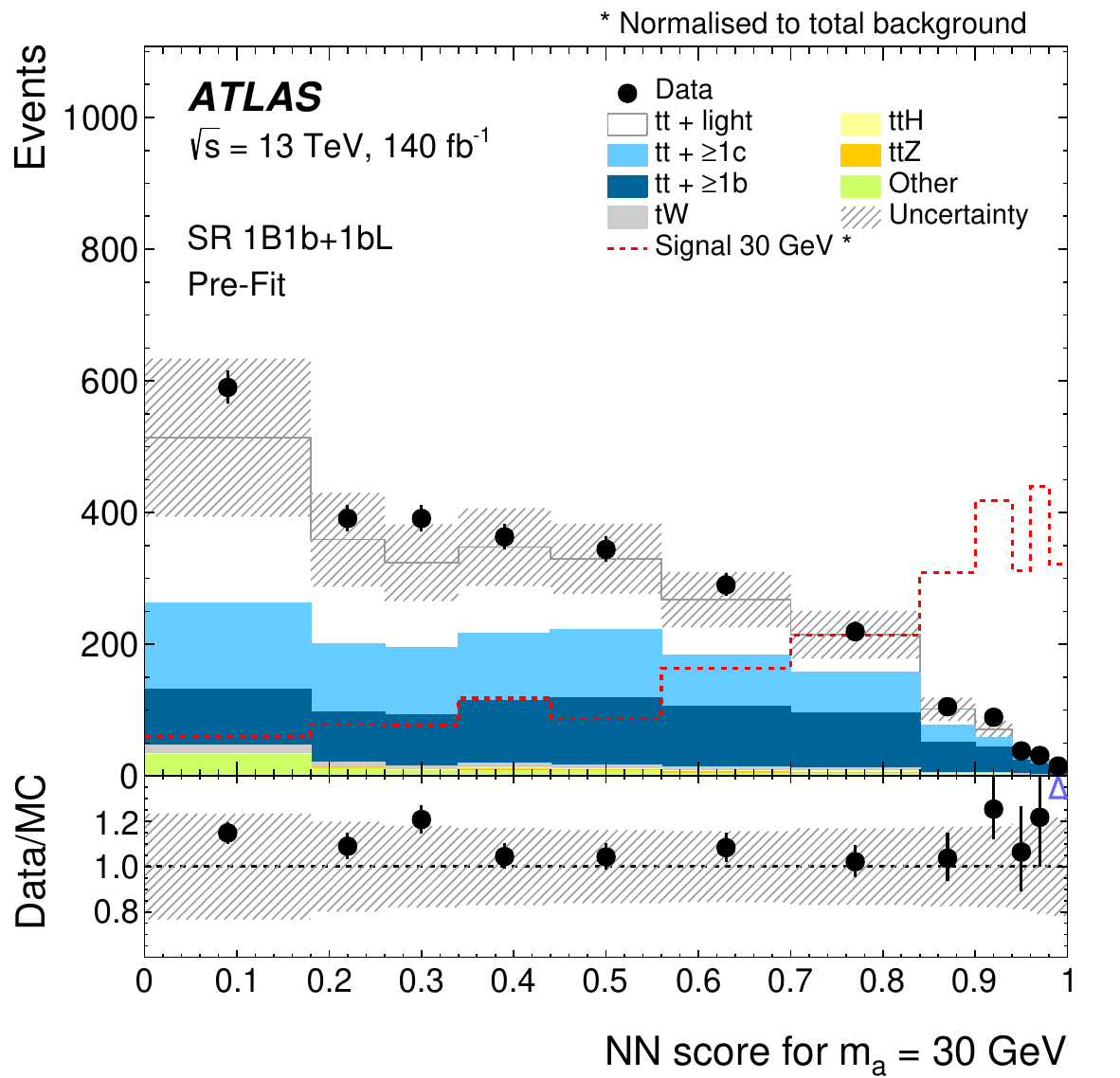}}
\caption{Pre-fit distributions corresponding to the NN output score of (a) SR 0B4b, (b) SR 0B3b, (c) SR 1B2b and (d) SR 1B1b+1bL for the 30~\GeV mass hypothesis fit. The dashed line shows the distribution of signal normalised to the total number of events in each region. The band displays the total pre-fit uncertainty. Arrows appearing in the bottom panels indicate the ratio being outside the displayed range.}
\label{fig:sec_Asimov_tta_30_NNscore_prefit}
\end{figure}


\section{Statistical treatment}
\label{ch:statistical_treatment}

To test for the presence of a $\tta$ signal, for each mass hypothesis, a binned maximum-likelihood fit to the data is performed simultaneously in all SRs and the CR (Section~\ref{sec:signal_control_regions}). In each SR, the input to the fit is the corresponding NN distribution described in Section~\ref{ch:NN}, evaluated at the appropriate mass hypothesis. In the CR, the input to the fit is the \texttt{sumPCBTag} distribution.
The parameter of interest is the signal strength, $\mu$, a multiplicative factor to the cross section of the signal process.
In addition to the signal strength $\mu$, the fit includes three additional free parameters that work as scale factors to the normalisation for the three main background components: $k$($\ttbar$+light, $tW$), $k$($\ttbar$+$\geq$1$c$), and $k$($\ttbar$+$\geq$1$b$). To estimate the signal strength, a binned likelihood function $\mathcal{L}(\mu,\theta)$ is used,

\begin{equation*}
\mathcal{L}(\mu,\theta) = \prod_i^N \frac{ (E[n_i(\mu,\theta)])^{n_i}}{n_i!}e^{-E[n_i(\mu,\theta)]}\prod_{\theta_j\in\theta}\rho(\theta_j|\tilde{\theta}_j).
\end{equation*}

The function is constructed as a product of Poisson probability terms with one Poisson term included for every bin $i$ of the NN distribution in the analysis regions. The binning of the NN distributions for each signal is chosen to provide a good separation of signal and background while maintaining a stable performance of the fit.
The expected number of events, $E[n_i(\mu,\theta)]$, in each bin, $n_i$, is a function of $\mu$, and a set of nuisance parameters, $\theta$.
The nuisance parameters encode effects from the normalisation of backgrounds, including the systematic uncertainties and one parameter per bin to model statistical uncertainties in the simulated samples. Unlike the free-floating parameters, all nuisance parameters are constrained by prior distributions, $\rho(\theta|\tilde{\theta})$, which follow Gaussian, log-normal, or Poisson distributions centred around their nominal values, $\tilde{\theta}$.
This procedure allows the reduction of the impact of the uncertainties by taking advantage of the separated populations of signal and background. The best-fit value of the signal strength is obtained by performing a fit to the data under the signal-plus-background hypothesis, maximising $\mathcal{L}(\mu,\theta)$ over $\mu$ and $\theta$.
To set upper limits on $\mu$, the following test statistic is used:

\begin{equation*}
\tilde{q}_\mu = \left\{
\begin{array}{ll}
-2 \ln \frac{ \mathcal{L}\left( \mu, \hat{\hat{\theta}}(\mu) \right) }{ \mathcal{L}\left( 0, \hat{\hat\theta}(0) \right) }  &  \hat\mu < 0,\\
-2 \ln \frac{ \mathcal{L} \left( \mu, \hat{\hat\theta}(\mu) \right) }{ \mathcal{L}\left( \hat\mu, \hat\theta \right) }  &  0\leq \hat\mu \leq \mu,\\
0 & \hat\mu > \mu.
\end{array}
\right.
\end{equation*}

The values of the signal strength and nuisance parameters that maximise the likelihood function are represented by $\hat\mu$ and $\hat\theta$, respectively.
For a given value of $\mu$, the values of the nuisance parameters that maximise the likelihood function are represented by $\hat{\hat\theta}(\mu)$. This test statistic measures the compatibility of the observed data with the background-only hypothesis ($\mu=0$), represented by the $p$-value, and is estimated by integrating the distribution of $\tilde{q}_0$ based on the asymptotic formula in Ref.~\cite{Cowan:2010js}. The test statistic is set to zero for $\hat\mu > \mu$, as this case indicates that the $\mu$ hypothesis is compatible with the observed data and cannot be rejected. Upper limits on $\mu$ are derived by using $\tilde{q}_\mu$ in the CL$_{\text{s}}$ method~\cite{Junk:1999kv,Read:2002hq}.

The systematic uncertainties, including those derived from MC samples, can show fluctuations due to generator weights or statistical variations. To ensure the quality of the templates and the stability of the fit, smoothing algorithms are applied to the histograms before the fit. In addition, systematic uncertainties are pruned to reduce computing time. Only uncertainties with an effect greater than 1\% are included in the fit. This is done separately for shape and normalisation effects.


\section{Systematic uncertainties}
\label{ch:systematic_uncertainties}

Various sources of systematic uncertainties are considered.
Each systematic uncertainty is introduced as a nuisance parameter (NP) in the statistical analysis described in Section~\ref{ch:statistical_treatment}. Section~\ref{sec:experimental_uncertainties} describes all experimental uncertainties, related to the luminosity and pile-up or the reconstruction and identification of jets and leptons. They are applied to all MC samples equally and their effects are treated in a correlated way across all four SRs and the CR in the final fit. The signal and background modelling uncertainties are detailed in Section~\ref{sec:modelling_uncertainties}, and can be different depending on the process.
They are implemented as decorrelated between regions, given their different coverage of phase spaces, and decorrelated between signal and background samples in the fit.

\subsection{Experimental uncertainties}
\label{sec:experimental_uncertainties}

\textbf{Luminosity and pile-up modelling.} The uncertainty in the integrated luminosity for the full Run 2 data sample is 0.83\%~\cite{DAPR-2021-01}, obtained using the LUCID-2 detector~\cite{LUCID2} for the primary luminosity measurements.
A variation in the pile-up reweighting of simulated events is included to cover the uncertainty in the ratio of the simulated and measured distribution of inelastic cross sections.

\textbf{Leptons.} Uncertainties associated with leptons are related to the trigger, reconstruction, identification and isolation, as well as the lepton energy or momentum scale and resolution.
The reconstruction, identification, and isolation efficiency of electrons and muons, as well as the efficiency of the trigger used to record the events, differ slightly between data and simulation, and is corrected by dedicated scale factors. Efficiency scale factors are measured using tag-and-probe techniques on $Z\rightarrow ll$ data and simulated samples~\cite{EGAM-2018-01,MUON-2018-03}, and are applied to the simulation to correct for differences.
Additional sources of uncertainty originate from the corrections applied to adjust the lepton momentum scale and resolution in the simulation to match those in data, measured using $Z\rightarrow ll$ and $J/\psi\rightarrow ll$ events~\cite{EGAM-2018-01,MUON-2018-03}.

\textbf{Jets.} Uncertainties associated with jets arise from the efficiency of pile-up rejection by the jet vertex tagger (JVT), from the jet energy scale (JES) and resolution (JER), and from the different flavour-tagging algorithms used, DL1r and DeXTer.
Scale factors are applied to correct for discrepancies between data and MC for JVT efficiencies, and are estimated by using $Z\rightarrow \mu\mu$ with tag-and-probe techniques~\cite{PERF-2014-03}.
The jet energy scale and its uncertainty are derived by combining information from test-beam data, LHC collision data and simulation~\cite{JETM-2018-05}.
The jet energy resolution is measured in Run 2 data and simulation as a function of jet $p_\text{T}$ and rapidity using dijet events.

To correct flavour-tagging efficiencies in simulated samples to match those measured in data, scale factors are derived. %
They are calculated as a function of $p_\text{T}$ for $b$-jets, $c$-jets, and light jets separately in dedicated calibration analyses. For $b$-jet efficiencies, $\ttbar$ events in the dilepton topology are used, exploiting the very pure sample of $b$-jets arising from the decay of the top quarks~\cite{FTAG-2018-01}. For $c$-jet mistag rates, $\ttbar$ events in the single-lepton topology are used, exploiting $c$-jets from the hadronically decaying $W$ boson~\cite{FTAG-2020-08}. The negative-tag method is used in $Z$+jets events~\cite{FTAG-2019-02} for light-jets mistag rates. The use of DeXTer introduces additional scale factors to correct for the differences in efficiency between simulated samples and data.
The scale factors for DeXTer are derived as a function of $p_\text{T}$ for $B$- and $b$-jets. The calibration measurements with data are performed using both $\ttbar$ and $Z$+jets events simultaneously to measure $B$-jet tagging and $b$-jet mistagging efficiency in data. Nevertheless, the DeXTer uncertainties are provided with conservative error bands, leaving the calibration to be performed in situ in the final fit of the analysis.
Further details on the methodology can be found in Ref.~\cite{ATL-PHYS-PUB-2022-042,ATL-PHYS-PUB-2022-042}.

\textbf{Missing transverse momentum.} All the described uncertainties in energy scales or resolutions of the reconstructed objects (hard components) are propagated to the missing transverse momentum. Additional uncertainties in the scale and resolution of the soft term are considered, to account for the disagreement between data and MC for the $p_\text{T}$ balance between the hard and soft components~\cite{JETM-2020-03}.

\textbf{Tracks.} Systematic uncertainties related to the track selection efficiency are determined by changing the amount of tracker material and the physical models in the \GEANT simulation~\cite{PERF-2015-08,IDTR-2019-05}.
Dedicated uncertainties are considered for the track parameters, including the transverse and longitudinal impact parameters and the track sagitta.

\textbf{Large-$R$ jet mass scale correction.} To correct for the mismodelling in the large-$R$ jet mass, additional mass scale corrections are estimated.
The large-$R$ jet mass scale is varied by $\pm$5\% and compared with the nominal results.

\subsection{Modelling uncertainties}
\label{sec:modelling_uncertainties}

\textbf{Renormalisation ($\muR$) and factorisation ($\muF$) scales.} Variations in the renormalisation and factorisation scales are used to estimate the uncertainty due to missing higher order corrections.
The uncertainties are combined by taking an envelope of all the variations.

\textbf{Initial-state radiation and final-state radiation modelling.}
For the ISR, the amount of radiation is increased (decreased) using the \texttt{Var3cUp} (\texttt{Var3cDown}) variation of the A14 tune~\cite{ATL-PHYS-PUB-2014-021}.
For the FSR, the amount of radiation is increased (decreased) varying the coupling of the QCD emission in the final state by a factor of 0.5 (2).

\textbf{PDF uncertainties.} The PDF uncertainties follow the PDF4LHC recommendations~\cite{Butterworth:2015oua}.
The $\alpha_\text{s}$ uncertainty is derived using the same PDF set evaluated with two different $\alpha_\text{s}$ values. The uncertainties from the PDF and $\alpha_\text{s}$ are added in quadrature.

\textbf{Parton shower.} The uncertainty associated with hadronisation and parton shower is evaluated by comparing samples with different parton shower models.
The nominal $\tta$ samples simulated using \POWPY[8] are compared with samples simulated using \MGNLOHER[7]~\cite{Bellm:2015jjp}. The comparison is done after normalising both \tta samples.
The nominal \ttbar (4FS and 5FS) \POWPY[8] samples are compared with samples simulated using \POWHER[7].
The $tW$ and $\ttH$ \MGNLOPY[8] samples are compared with \MGNLOHER[7] samples.

\textbf{Matrix element uncertainties.} For the 5FS and 4FS \ttbar samples, the uncertainty associated with the matching between the Matrix element calculations and the parton shower is calculated by comparing the nominal \POWPY[8] sample with an alternative set of samples simulated also in \POWPY[8] but using the \texttt{pThard=1} setting. For $tW$ and $\ttH$, the matrix element uncertainty is evaluated by comparing the nominal \POWPY[8] samples to those simulated with \MGNLOPY[8]. For $\ttZ$, the nominal samples are compared with an alternative sample simulated using \SHERPA[2.2.0], which accounts both for the matrix element and parton shower uncertainties.

\textbf{\POWHEG damping function.} In the $t\bar{t}b\bar{b}$ (4FS) samples, the effect of the choice of a damping scale $h_{\mathrm{bzd}}$ that controls the resummation of infrared divergences is evaluated by comparing the nominal sample ($h_\mathrm{bzd}=5$) with an alternative sample in which the scale is set to 2~\cite{ATL-PHYS-PUB-2022-006}.

\textbf{Initial-state shower recoil.} The uncertainty due to the recoil choice of ISR emissions is evaluated by comparing the nominal sample, in which the whole final state recoils the ISR emission, with an alternative one, in which only one final-state parton recoils against the ISR emission~\cite{ATL-PHYS-PUB-2022-006}.

\textbf{Interference between \ttbar and $tW$.} To account for uncertainties in the interference between $\ttbar$+jets and $tW$, the nominal $tW$ sample simulated using diagram removal (DR) is compared with another sample simulated using diagram subtraction (DS).

\textbf{Reweighting uncertainties.} To account for the systematic uncertainties associated with the reweighting functions described in Section~\ref{ch:reweighting}, several uncertainties are determined by the variations of $\ttbar$+light + $tW$, $\ttbar$+$\geq$1$c$ and $\ttbar$+$\geq$1$b$ normalisation factors and the variations of the parameters of the $H_\text{T}^\text{red}$ hyperbolic fit. The uncertainties are evaluated after diagonalising the fit correlation matrix and propagating the diagonal variations in a correlated way.


\section{Results}
\label{ch:results}

The expected and observed upper limits on the inclusive $\sigma(\tta)\times \text{BR}(a\rightarrow\bbbar)$ are shown in Figure~\ref{fig:sec_SplusB_expectedLimit} as a function of the $a$-boson mass, which ranges from 12 to 100~\GeV. This result is compared with the predicted cross sections for the signal corresponding to three different values of the coupling of the $a$-boson to the top quark, defined as a strength modifier to the SM Yukawa coupling. No significant excess is observed: the largest excess corresponds to the 30~\GeV mass hypothesis, with a local significance of 2.0 standard deviations. Assuming $\text{BR}(a\rightarrow\bbbar)=100\%$, the mass region between 50 and 80~\GeV is excluded for a coupling of the pseudoscalar to the top quark of 0.5, while a coupling of 1.0 is excluded for all masses.
Post-fit distributions of the NN output score corresponding to this mass in each of the four signal regions and of the sum of the pseudo-continuous $b$-tagging score of all jets in the control region are shown in Figure~\ref{fig:SplusB_tta_30_postfit}. Table~\ref{tab:event_yields_SRsCR_30postfit} shows the post-fit event yields per signal and background component in each of the signal and control regions for the same mass hypothesis.

\begin{table}[tb]
\caption{Post-fit background and signal yields in the four signal regions and in the control region for the $m_{a} = 30$~\GeV hypothesis. %
The uncertainties in each yield are the total uncertainties of each component after the fit.}
\centering
\begin{tabular}{l|r@{}c@{}l|r@{}c@{}l|r@{}c@{}l|r@{}c@{}l|r@{}c@{}l}

Sample & \multicolumn{3}{c|}{SR 0B4b} & \multicolumn{3}{c|}{SR 0B3b} & \multicolumn{3}{c|}{SR 1B2b} & \multicolumn{3}{c|}{SR 1B1b+1bL} & \multicolumn{3}{c}{CR 0B2b+1bL} \\ \hline
\hline
Signal 30~\GeV
& 28  & $\;\pm\;$ & 14
& 180 & $\;\pm\;$ & 98
& 71  & $\;\pm\;$ & 31
& 35  & $\;\pm\;$ & 16
& 110 & $\;\pm\;$ & 63 \\
$\ttbar$+light
& 5     & $\;\pm\;$ & 2
& 1400  & $\;\pm\;$ & 320
& 130   & $\;\pm\;$ & 24
& 1100  & $\;\pm\;$ & 180
& 17000 & $\;\pm\;$ & 3000 \\
$\ttbar$+$\geq$1$c$
& 58     & $\;\pm\;$ & 21
& 4700   & $\;\pm\;$ & 1200
& 380    & $\;\pm\;$ & 140
& 740    & $\;\pm\;$ & 230
& 12000  & $\;\pm\;$ & 3500 \\
$\ttbar$+$\geq$1$b$
& 1093 & $\;\pm\;$ & 47
& 9820 & $\;\pm\;$ & 700
& 1758 & $\;\pm\;$ & 97
& 815  & $\;\pm\;$ & 70
& 5510 & $\;\pm\;$ & 650\\
$tW$
& 22  & $\;\pm\;$ & 12
& 360 & $\;\pm\;$ & 140
& 46  & $\;\pm\;$ & 20
& 64  & $\;\pm\;$ & 17
& 830 & $\;\pm\;$ & 220 \\
$\ttbar H$
& 62  & $\;\pm\;$ & 9
& 222 & $\;\pm\;$ & 22
& 31  & $\;\pm\;$ & 4
& 14  & $\;\pm\;$ & 12
& 136 & $\;\pm\;$ & 13 \\
$\ttbar Z$
& 27  & $\;\pm\;$ & 6
& 120 & $\;\pm\;$ & 22
& 15  & $\;\pm\;$ & 3
& 11  & $\;\pm\;$ & 2
& 128 & $\;\pm\;$ & 25 \\
Other
& 14   & $\;\pm\;$ & 2
& 394  & $\;\pm\;$ & 35
& 47   & $\;\pm\;$ & 4
& 78   & $\;\pm\;$ & 10
& 1060 & $\;\pm\;$ & 120 \\ \hline
Total pred.
& 1300  & $\;\pm\;$ & 35
& 17000 & $\;\pm\;$ & 130
& 2500  & $\;\pm\;$ & 50
& 2900  & $\;\pm\;$ & 53
& 36000 & $\;\pm\;$ & 190 \\
Data
& 1301  &&
& 17242 &&
& 2479  &&
& 2866  &&
& 36350 && \\

\multicolumn{2}{c}{ }\\
\end{tabular}

\label{tab:event_yields_SRsCR_30postfit}
\end{table}

\begin{figure}[bt]
\centering
\includegraphics[width=0.6\textwidth]{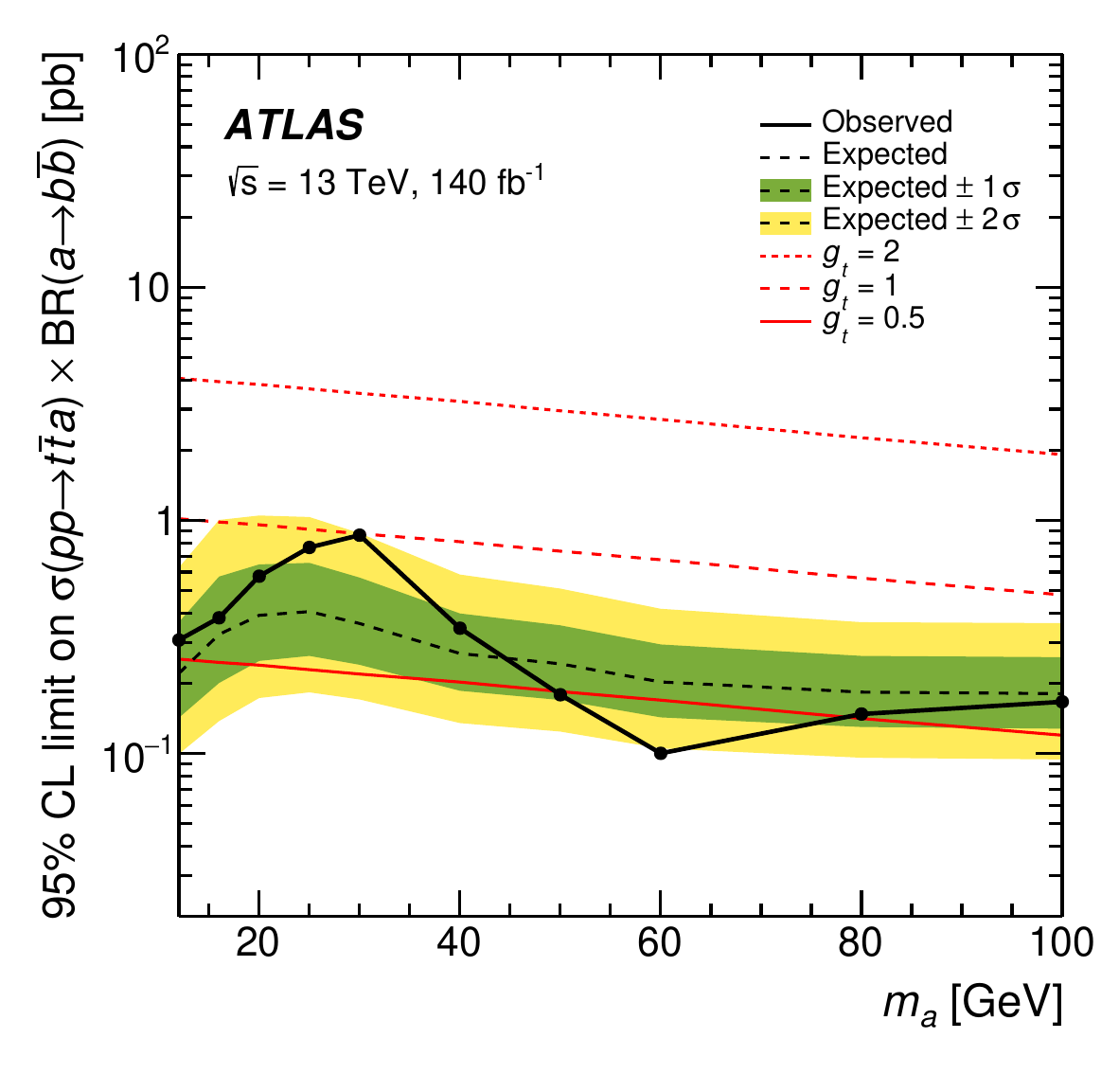}
\caption{Expected and observed 95\% CL upper limits of $\sigma(\tta)\times \text{BR}(a\rightarrow\bbbar)$ as a function of the $a$-boson mass. The lines correspond to the signal cross sections calculated using different coupling strengths of the $a$ boson to the top quark assuming a $\text{BR}(a\rightarrow\bbbar)=100\%$.}
\label{fig:sec_SplusB_expectedLimit}
\end{figure}

\begin{figure}[tb]
\centering

\subfloat[]{\includegraphics[width=0.4\textwidth]{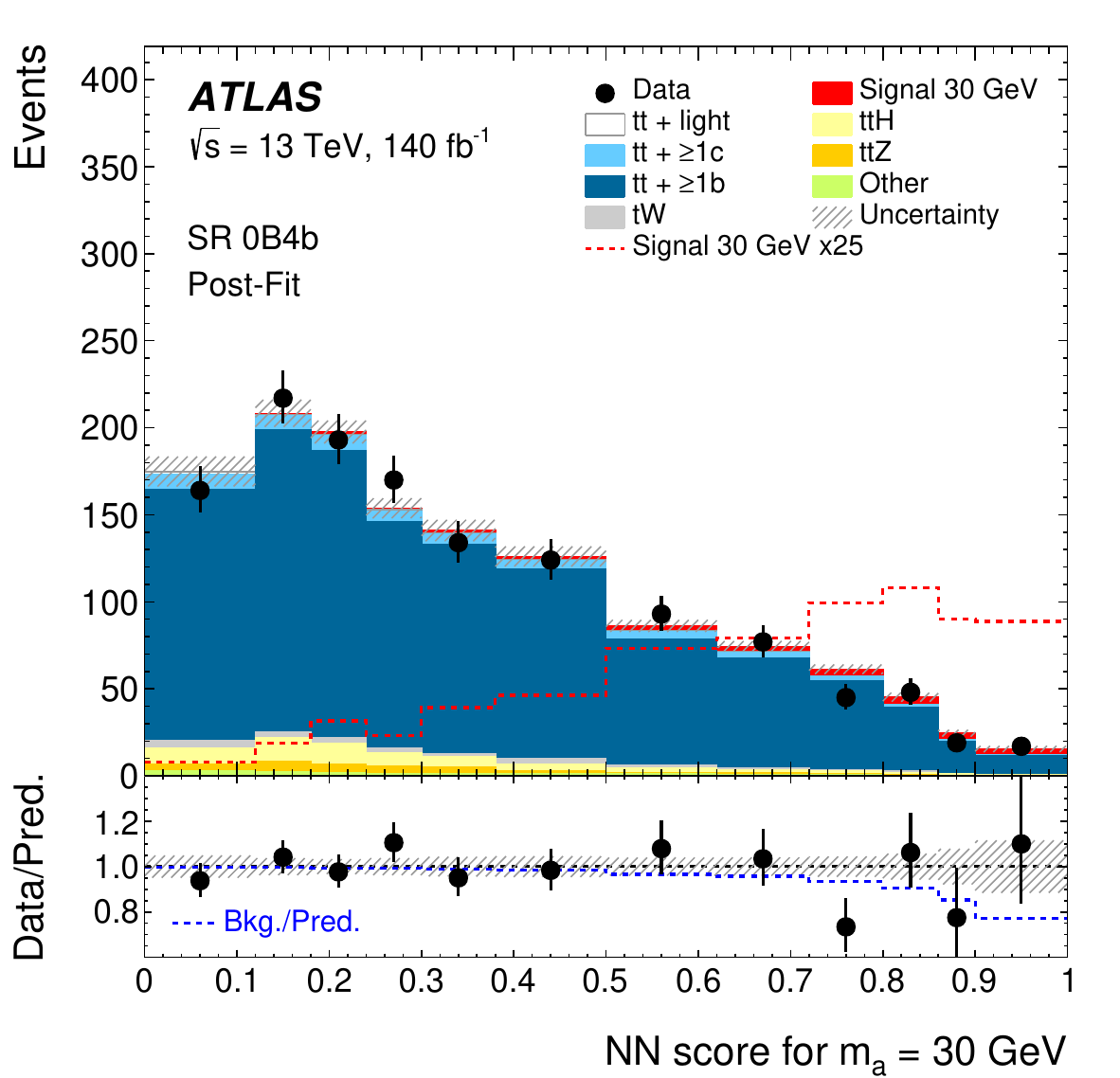}}
\subfloat[]{\includegraphics[width=0.4\textwidth]{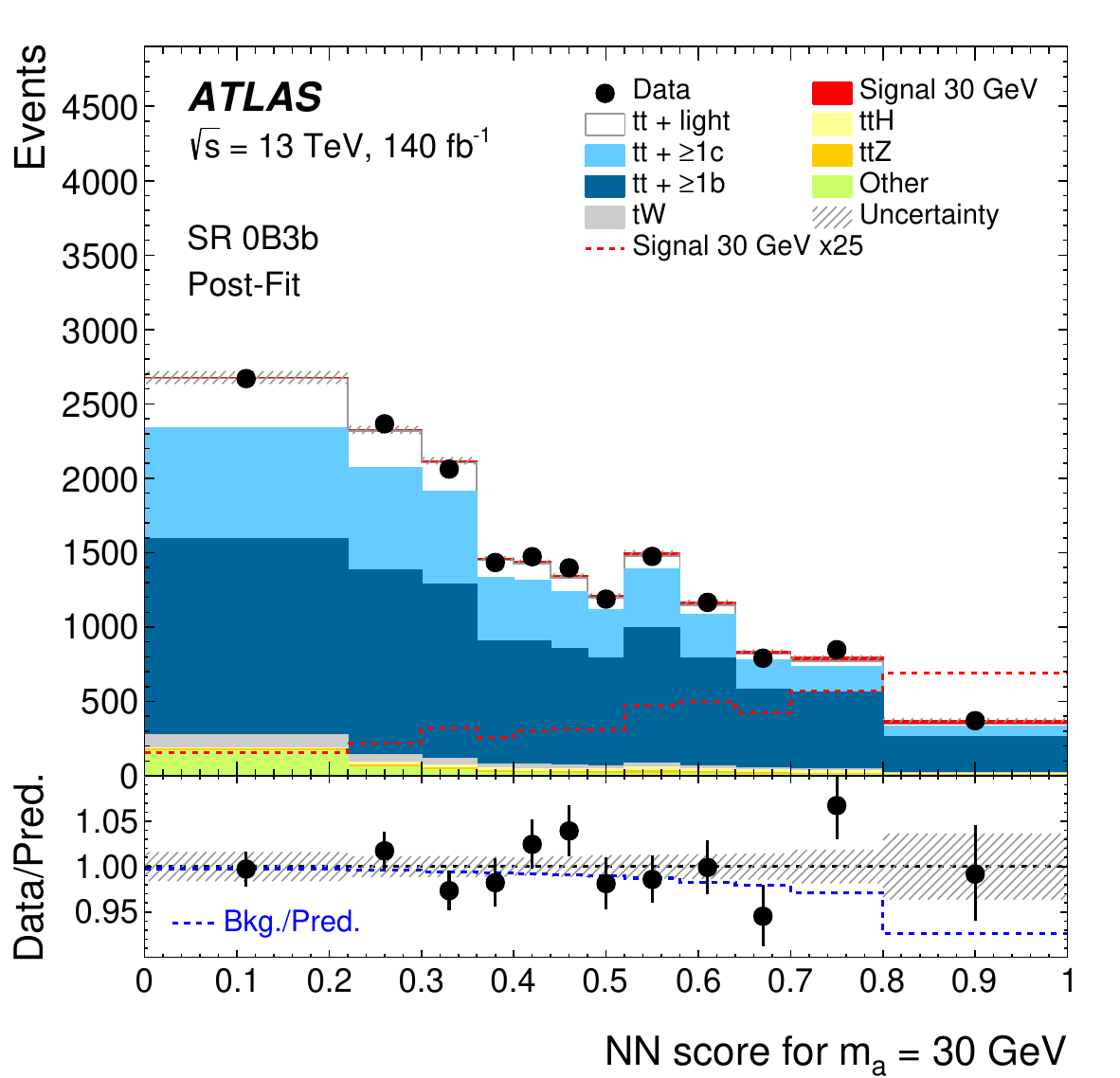}}\\
\subfloat[]{\includegraphics[width=0.4\textwidth]{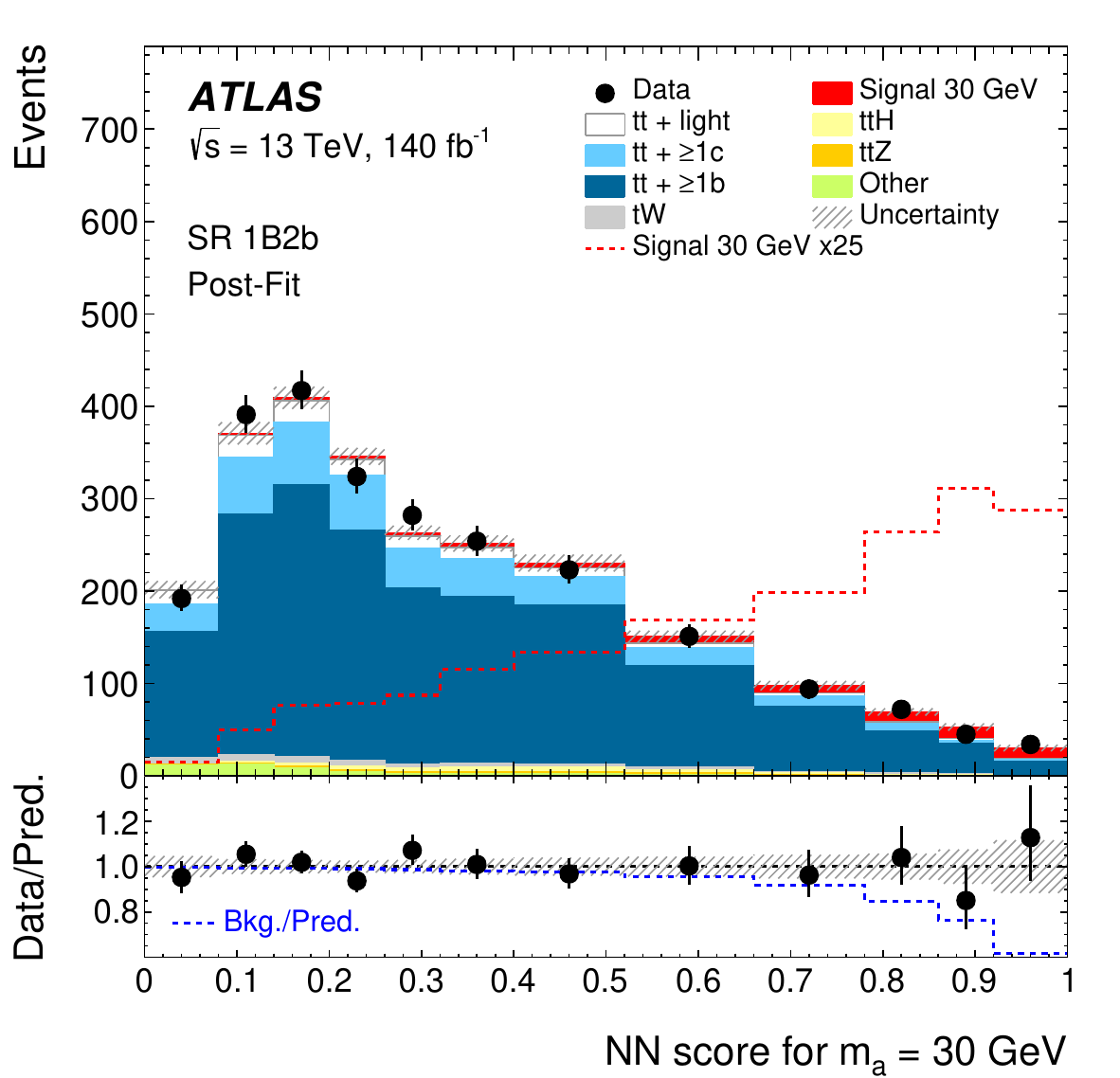}}
\subfloat[]{\includegraphics[width=0.4\textwidth]{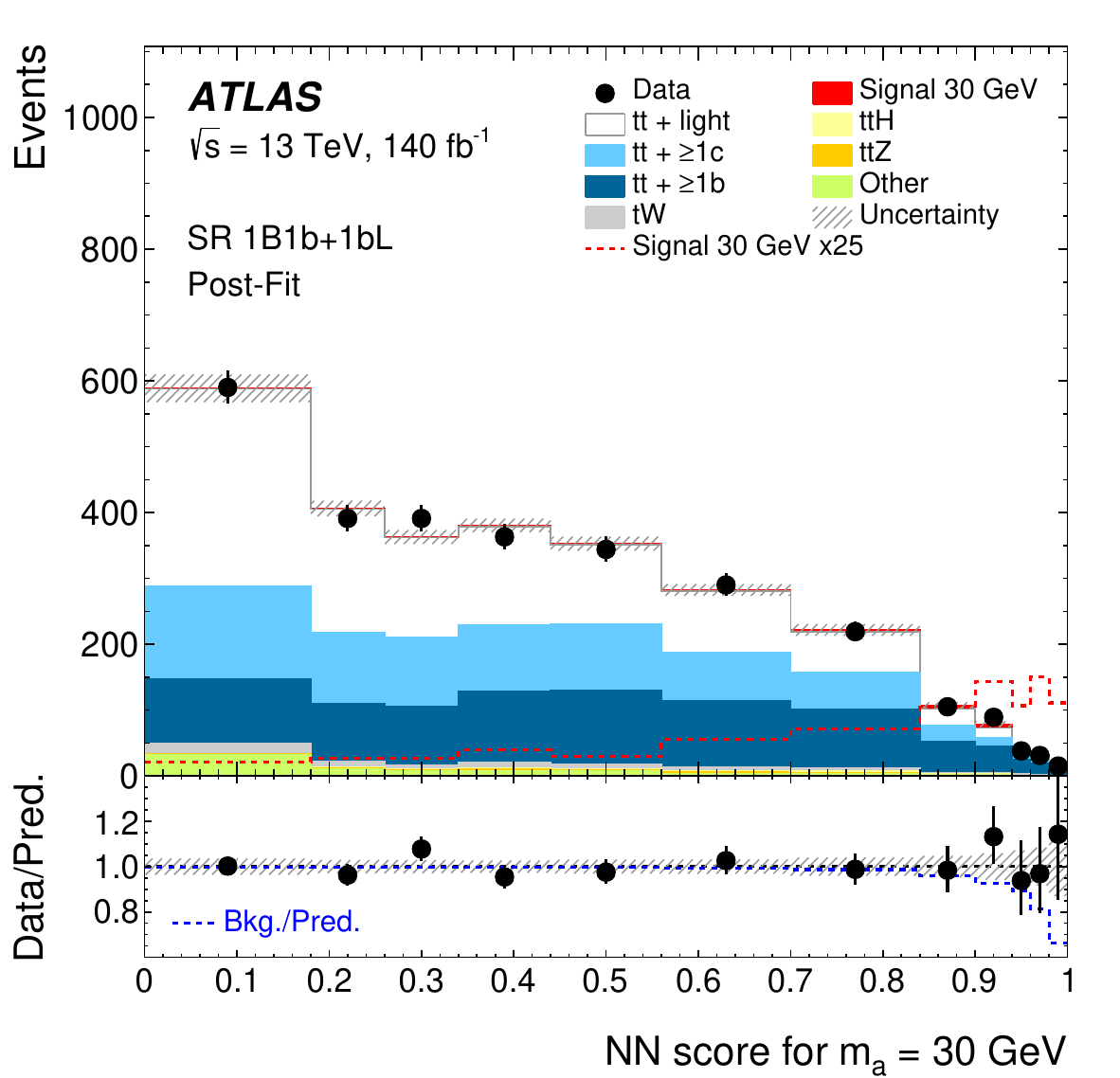}}\\
\subfloat[]{\includegraphics[width=0.4\textwidth]{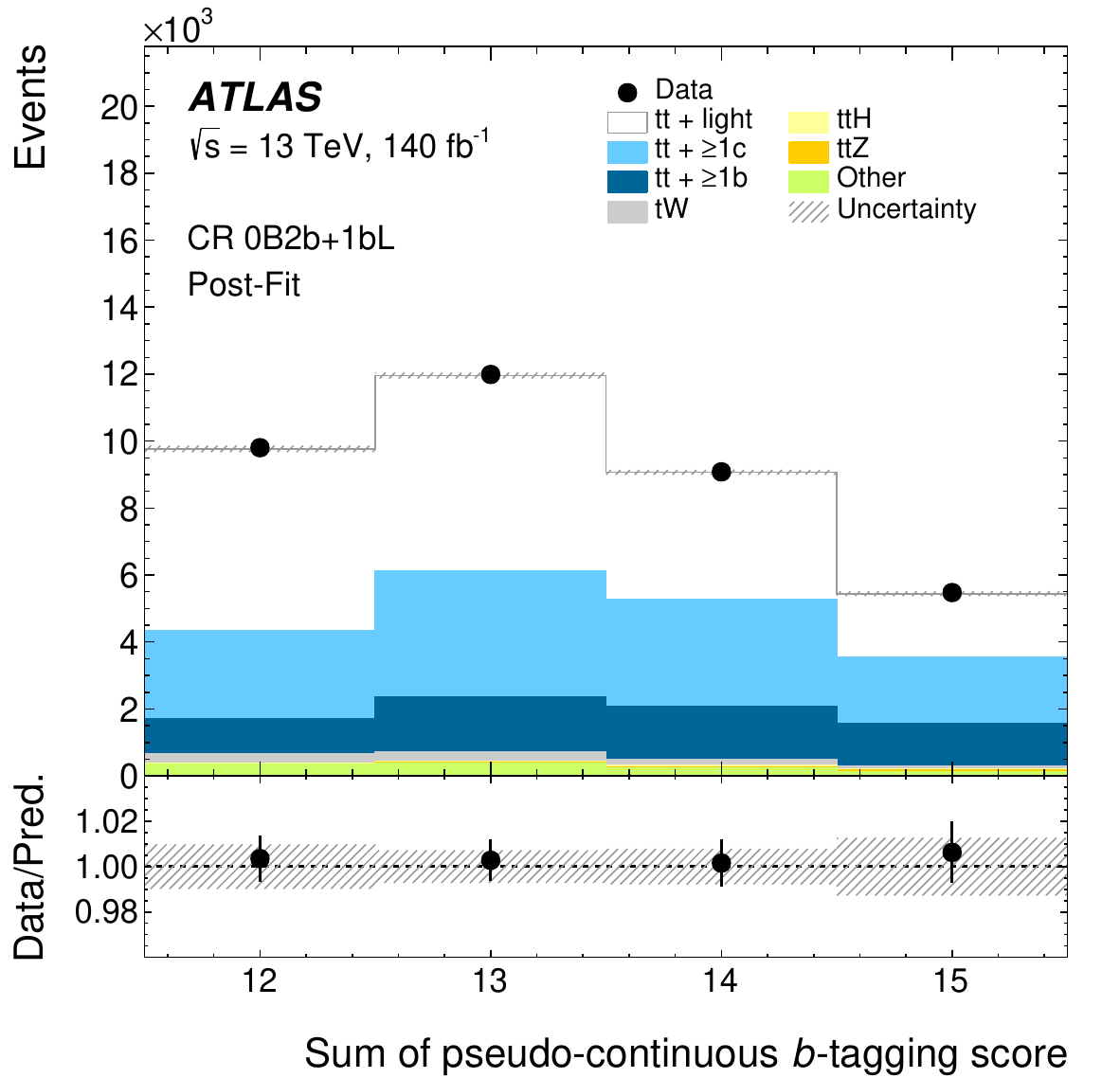}}

\caption{Post-fit distributions corresponding to the NN output score of (a) SR 0B4b, (b) SR 0B3b, (c) SR 1B2b and (d) SR 1B1b+1bL and to the sum of the pseudo-continuous $b$-tagging score of (e) CR 0B2b+1bL for the 30~\GeV mass hypothesis fit. The dashed lines in the top and ratio panels show the post-fit distribution of the signal scaled by a factor of 25 and the post-fit ratio of the background over the total prediction, respectively. The band displays the total post-fit uncertainty.}
\label{fig:SplusB_tta_30_postfit}
\end{figure}

Table~\ref{tab:ImpactTable30} summarises the impact of the different sources of uncertainties in the fitted signal strength for three different mass hypotheses: 12, 30 and 80~\GeV, which are representative of the low, medium and high mass ranges, respectively.
Fits to low-mass hypotheses are limited by data statistics, track reconstruction and \DeXTer-related uncertainties.
Fits to medium-mass hypotheses are dominated by \DeXTer-related uncertainties, followed by the modelling of the $\ttbar$+$\geq$1$b$ process and data statistics. Finally, fits to high-mass hypotheses are limited mainly by data statistics, the modelling of the $\ttbar$+$\geq$1$b$ process and the normalisation of $\ttbar$+HF\@.
In all cases, the uncertainties in the modelling of the signal are subdominant compared with that of $\ttbar$+$\geq$1$b$. No large pulls are observed in any of the fits.
Including the pre-fit reweighting corrections detailed in Table~\ref{tab:Correction1_ttbarWt_NormFactors}, the final normalisation factors extracted in the fit corresponding to the 30~GeV mass hypothesis are $1.0\pm0.3$  for $\ttbar$+light and $tW$, $1.5\pm0.5$ for $\ttbar$+$\geq$1$c$ and $1.2\pm0.2$ for $\ttbar$+$\geq$1$b$. These results are compatible with the latest ATLAS $\ttbar H$ Run 2 analysis~\cite{HIGG-2020-24}.

\begin{table}[tb]
\caption{Table of the impact of each group of uncertainties in the fitted cross section for the hypothesis masses of 12, 30 and 80~\GeV. The values shown are the average of up and down uncertainties. The fitted cross section values include the BR($t\bar{t}\rightarrow Wb Wb$) $\times$ BR($W\to l\nu$) $\times$ BR($W\to l \nu$) in addition to the BR($a\rightarrow b\bar{b}$).}

\centering
\begin{tabular}{l|c|c|c}

& $m_{a}=12$~GeV & $m_{a}=30$~GeV & $m_{a}=80$~GeV \\
Fitted cross section [fb] &$\hat{\sigma}=9$&$\hat{\sigma}=46$&$\hat{\sigma}=-6.1$\\
\hline
\hline
Uncertainty source & $\Delta\hat{\sigma}$ & $\Delta\hat{\sigma}$ & $\Delta\hat{\sigma}$  \\ \hline
Data statistics & 6.1 & 11.0 & 6.0 \\
MC statistics & 2.4 & 4.2 & 1.8 \\
\hline
Luminosity \& pile-up & 0.1 & 0.4 & 0.1 \\
Jet reconstruction & 0.5 & 4.9 & 1.2 \\
Lepton reconstruction & <0.1 & <0.1 & <0.1 \\
\met reconstruction & <0.1 & 0.3 & <0.1 \\
Track reconstruction & 4.1 & 1.5 & 0.1 \\
DL1r & 0.4 & 3.5 & 1.4 \\
DeXTer & 4.2 & 18 & 1.1 \\
\hline
Modelling signal & 1.7 & 7.5 & 1.3 \\
Modelling $\ttbar+b$ & 2.7 & 13 & 5.5 \\
Modelling $\ttbar+c$ & 0.9 & 1.8 & 1.4 \\
Modelling $\ttbar+$light & 0.8 & 2.0 & 2.2 \\
Modelling tW & 0.3 & 0.7 & 0.6 \\
Modelling ttH & 0.1 & 0.3 & 0.2 \\
Modelling ttZ & 0.1 & 0.2 & 1.0 \\
\hline
Norm factors & 0.7 & 6.7 & 4.7 \\
Reweighting & <0.1 & <0.1 & <0.1 \\
\hline
Total systematic uncertainty & 8.0 & 22 & 7.8 \\
Total uncertainty & 10 & 24 & 9.7 \\

\multicolumn{2}{c}{ }\\
\end{tabular}

\label{tab:ImpactTable30}
\end{table}


\clearpage
\section{Conclusions}
\label{ch:conclusions}

A search for a pseudoscalar $a$ produced in association with either a pair of top quarks or a single top and a $W$ boson in the dilepton decay channel is performed using the
full \RunTwo $pp$ data sample collected by the ATLAS detector at the LHC. The search targets the dominant decay channel of the pseudoscalar mass probed in this analysis: $a\rightarrow b\bar{b}$. The search covers the pseudoscalar boson mass between 12 and 100~\GeV, involving both the kinematic regime where the decay products of the pseudoscalar merge into large $B$-jets and the regime where the $b$-tagged jets are resolved. Limits on the signal production cross section times the branching ratio of the decay into a pair of bottom quarks are extracted. Assuming $\text{BR}(a\rightarrow\bbbar)=100\%$, the mass region between 50 and 80~\GeV is excluded for a coupling of the pseudoscalar to the top quark of 0.5, while a coupling of 1.0 is excluded for all masses. These model independent results are the first limits of their kind and complement previous searches by ATLAS~\cite{HDBS-2020-12} and CMS~\cite{CMS-EXO-21-018} exploring leptonic decays of the pseudoscalar.


%
%

\section*{Acknowledgements}

%
%

%
%

We thank CERN for the very successful operation of the LHC and its injectors, as well as the support staff at
CERN and at our institutions worldwide without whom ATLAS could not be operated efficiently.

The crucial computing support from all WLCG partners is acknowledged gratefully, in particular from CERN, the ATLAS Tier-1 facilities at TRIUMF/SFU (Canada), NDGF (Denmark, Norway, Sweden), CC-IN2P3 (France), KIT/GridKA (Germany), INFN-CNAF (Italy), NL-T1 (Netherlands), PIC (Spain), RAL (UK) and BNL (USA), the Tier-2 facilities worldwide and large non-WLCG resource providers. Major contributors of computing resources are listed in Ref.~\cite{ATL-SOFT-PUB-2025-001}.

We gratefully acknowledge the support of ANPCyT, Argentina; YerPhI, Armenia; ARC, Australia; BMWFW and FWF, Austria; ANAS, Azerbaijan; CNPq and FAPESP, Brazil; NSERC, NRC and CFI, Canada; CERN; ANID, Chile; CAS, MOST and NSFC, China; Minciencias, Colombia; MEYS CR, Czech Republic; DNRF and DNSRC, Denmark; IN2P3-CNRS and CEA-DRF/IRFU, France; SRNSFG, Georgia; BMBF, HGF and MPG, Germany; GSRI, Greece; RGC and Hong Kong SAR, China; ICHEP and Academy of Sciences and Humanities, Israel; INFN, Italy; MEXT and JSPS, Japan; CNRST, Morocco; NWO, Netherlands; RCN, Norway; MNiSW, Poland; FCT, Portugal; MNE/IFA, Romania; MSTDI, Serbia; MSSR, Slovakia; ARIS and MVZI, Slovenia; DSI/NRF, South Africa; MICIU/AEI, Spain; SRC and Wallenberg Foundation, Sweden; SERI, SNSF and Cantons of Bern and Geneva, Switzerland; NSTC, Taipei; TENMAK, T\"urkiye; STFC/UKRI, United Kingdom; DOE and NSF, United States of America.

Individual groups and members have received support from BCKDF, CANARIE, CRC and DRAC, Canada; CERN-CZ, FORTE and PRIMUS, Czech Republic; COST, ERC, ERDF, Horizon 2020, ICSC-NextGenerationEU and Marie Sk{\l}odowska-Curie Actions, European Union; Investissements d'Avenir Labex, Investissements d'Avenir Idex and ANR, France; DFG and AvH Foundation, Germany; Herakleitos, Thales and Aristeia programmes co-financed by EU-ESF and the Greek NSRF, Greece; BSF-NSF and MINERVA, Israel; NCN and NAWA, Poland; La Caixa Banking Foundation, CERCA Programme Generalitat de Catalunya and PROMETEO and GenT Programmes Generalitat Valenciana, Spain; G\"{o}ran Gustafssons Stiftelse, Sweden; The Royal Society and Leverhulme Trust, United Kingdom.

In addition, individual members wish to acknowledge support from Armenia: Yerevan Physics Institute (FAPERJ); CERN: European Organization for Nuclear Research (CERN DOCT); Chile: Agencia Nacional de Investigaci\'on y Desarrollo (FONDECYT 1230812, FONDECYT 1230987, FONDECYT 1240864); China: Chinese Ministry of Science and Technology (MOST-2023YFA1605700, MOST-2023YFA1609300), National Natural Science Foundation of China (NSFC - 12175119, NSFC 12275265); Czech Republic: Czech Science Foundation (GACR - 24-11373S), Ministry of Education Youth and Sports (ERC-CZ-LL2327, FORTE CZ.02.01.01/00/22\_008/0004632), PRIMUS Research Programme (PRIMUS/21/SCI/017); EU: H2020 European Research Council (ERC - 101002463); European Union: European Research Council (ERC - 948254, ERC 101089007, ERC, BARD, 101116429), European Regional Development Fund (SMASH COFUND 101081355, SLO ERDF), Horizon 2020 Framework Programme (MUCCA - CHIST-ERA-19-XAI-00), European Union, Future Artificial Intelligence Research (FAIR-NextGenerationEU PE00000013), Horizon 2020 (EuroHPC - EHPC-DEV-2024D11-051), Italian Center for High Performance Computing, Big Data and Quantum Computing (ICSC, NextGenerationEU); France: Agence Nationale de la Recherche (ANR-21-CE31-0022, ANR-22-EDIR-0002); Germany: Baden-Württemberg Stiftung (BW Stiftung-Postdoc Eliteprogramme), Deutsche Forschungsgemeinschaft (DFG - 469666862, DFG - CR 312/5-2); China: Research Grants Council (GRF); Italy: Istituto Nazionale di Fisica Nucleare (ICSC, NextGenerationEU), Ministero dell'Università e della Ricerca (NextGenEU I53D23000820006 M4C2.1.1); Japan: Japan Society for the Promotion of Science (JSPS KAKENHI JP22H01227, JSPS KAKENHI JP22H04944, JSPS KAKENHI JP22KK0227, JSPS KAKENHI JP23KK0245, JSPS KAKENHI JP24K23939); Norway: Research Council of Norway (RCN-314472); Poland: Ministry of Science and Higher Education (IDUB AGH, POB8, D4 no 9722), Polish National Science Centre (NCN 2021/42/E/ST2/00350, NCN OPUS 2023/51/B/ST2/02507, NCN OPUS nr 2022/47/B/ST2/03059, NCN UMO-2019/34/E/ST2/00393, UMO-2022/47/O/ST2/00148, UMO-2023/49/B/ST2/04085, UMO-2023/51/B/ST2/00920, UMO-2024/53/N/ST2/00869); Portugal: Foundation for Science and Technology (FCT); Spain: Generalitat Valenciana (Artemisa, FEDER, IDIFEDER/2018/048), Ministry of Science and Innovation (MCIN \& NextGenEU PCI2022-135018-2, MICIN \& FEDER PID2021-125273NB, RYC2019-028510-I, RYC2020-030254-I, RYC2021-031273-I, RYC2022-038164-I); Sweden: Carl Trygger Foundation (Carl Trygger Foundation CTS 22:2312), Swedish Research Council (Swedish Research Council 2023-04654, VR 2021-03651, VR 2022-03845, VR 2022-04683, VR 2023-03403, VR 2024-05451), Knut and Alice Wallenberg Foundation (KAW 2018.0458, KAW 2022.0358, KAW 2023.0366); Switzerland: Swiss National Science Foundation (SNSF - PCEFP2\_194658); United Kingdom: Leverhulme Trust (Leverhulme Trust RPG-2020-004), Royal Society (NIF-R1-231091); United States of America: U.S. Department of Energy (ECA DE-AC02-76SF00515), Neubauer Family Foundation.

%
%


\clearpage
\printbibliography
\clearpage
\input{atlas_authlist}
%

%
%

%

%
%
%
%

\end{document}

%% file: atlas_authlist.tex
 
\begin{flushleft}
\hypersetup{urlcolor=black}
{\Large The ATLAS Collaboration}

\bigskip

\AtlasOrcid[0000-0002-6665-4934]{G.~Aad}$^\textrm{\scriptsize 104}$,
\AtlasOrcid[0000-0001-7616-1554]{E.~Aakvaag}$^\textrm{\scriptsize 17}$,
\AtlasOrcid[0000-0002-5888-2734]{B.~Abbott}$^\textrm{\scriptsize 123}$,
\AtlasOrcid[0000-0002-0287-5869]{S.~Abdelhameed}$^\textrm{\scriptsize 119a}$,
\AtlasOrcid[0000-0002-1002-1652]{K.~Abeling}$^\textrm{\scriptsize 55}$,
\AtlasOrcid[0000-0001-5763-2760]{N.J.~Abicht}$^\textrm{\scriptsize 49}$,
\AtlasOrcid[0000-0002-8496-9294]{S.H.~Abidi}$^\textrm{\scriptsize 30}$,
\AtlasOrcid[0009-0003-6578-220X]{M.~Aboelela}$^\textrm{\scriptsize 45}$,
\AtlasOrcid[0000-0002-9987-2292]{A.~Aboulhorma}$^\textrm{\scriptsize 36e}$,
\AtlasOrcid[0000-0001-5329-6640]{H.~Abramowicz}$^\textrm{\scriptsize 157}$
% ...
\bigskip
\\

% ...
$^{17}$Department for Physics and Technology, University of Bergen, Bergen; Norway.\\
% ...
$^{30}$Physics Department, Brookhaven National Laboratory, Upton NY; United States of America.\\
% ...
$^{36}$$^{(a)}$Facult\'e des Sciences Ain Chock, Universit\'e Hassan II de Casablanca;$^{(b)}$Facult\'{e} des Sciences, Universit\'{e} Ibn-Tofail, K\'{e}nitra;$^{(c)}$Facult\'e des Sciences Semlalia, Universit\'e Cadi Ayyad, LPHEA-Marrakech;$^{(d)}$LPMR, Facult\'e des Sciences, Universit\'e Mohamed Premier, Oujda;$^{(e)}$Facult\'e des sciences, Universit\'e Mohammed V, Rabat;$^{(f)}$Institute of Applied Physics, Mohammed VI Polytechnic University, Ben Guerir; Morocco.\\
% ...
$^{45}$Physics Department, Southern Methodist University, Dallas TX; United States of America.\\
% ...
$^{49}$Fakult\"{a}t Physik , Technische Universit{\"a}t Dortmund, Dortmund; Germany.\\
% ...
$^{55}$II. Physikalisches Institut, Georg-August-Universit\"{a}t G\"ottingen, G\"ottingen; Germany.\\
% ...
$^{104}$CPPM, Aix-Marseille Universit\'e, CNRS/IN2P3, Marseille; France.\\
% ...
$^{119}$$^{(a)}$New York University Abu Dhabi, Abu Dhabi;$^{(b)}$United Arab Emirates University, Al Ain; United Arab Emirates.\\
% ...
$^{123}$Homer L. Dodge Department of Physics and Astronomy, University of Oklahoma, Norman OK; United States of America.\\
% ...
$^{157}$Raymond and Beverly Sackler School of Physics and Astronomy, Tel Aviv University, Tel Aviv; Israel.\\
% ...
$^{a}$ Also at Affiliated with an institute formerly covered by a cooperation agreement with CERN.\\
$^{b}$ Also at An-Najah National University, Nablus; Palestine.\\
$^{c}$ Also at Borough of Manhattan Community College, City University of New York, New York NY; United States of America.\\
$^{d}$ Also at Center for Interdisciplinary Research and Innovation (CIRI-AUTH), Thessaloniki; Greece.\\
$^{e}$ Also at Centre of Physics of the Universities of Minho and Porto (CF-UM-UP); Portugal.\\
$^{f}$ Also at CERN, Geneva; Switzerland.\\
% ...

\end{flushleft}
